\documentclass[english]{IEEEtran}
\usepackage[T1]{fontenc}
\usepackage[latin9]{inputenc}
\usepackage{color}
\usepackage{babel}
\usepackage{float}
\usepackage{url}
\usepackage{amsmath}
\usepackage{amsthm}
\usepackage{amssymb}
\usepackage{graphicx}
\usepackage{esint}
\usepackage[unicode=true,
 bookmarks=true,bookmarksnumbered=false,bookmarksopen=false,
 breaklinks=false,pdfborder={0 0 1},backref=false,colorlinks=false]
 {hyperref}
\hypersetup{pdftitle={Fast Numerical Nonlinear Fourier Transforms},
 pdfauthor={Sander Wahls and H. Vincent Poor},
 pdfkeywords={Nonlinear Fourier Transform, Forward Scattering Transform, Nonlinear Schroedinger Equation, Fast Algorithms}}

\makeatletter

\providecommand{\tabularnewline}{\\}
\floatstyle{ruled}
\newfloat{algorithm}{tbp}{loa}
\providecommand{\algorithmname}{Algorithm}
\floatname{algorithm}{\protect\algorithmname}

\theoremstyle{plain}
\newtheorem{thm}{\protect\theoremname}
\theoremstyle{remark}
\newtheorem{rem}[thm]{\protect\remarkname}
\theoremstyle{definition}
\newtheorem{example}[thm]{\protect\examplename}
\newenvironment{lyxlist}[1]
{\begin{list}{}
{\settowidth{\labelwidth}{#1}
 \setlength{\leftmargin}{\labelwidth}
 \addtolength{\leftmargin}{\labelsep}
 }}
{\end{list}}
\theoremstyle{plain}
\newtheorem{lem}[thm]{\protect\lemmaname}

\usepackage{amsmath}
\renewcommand\Re{\operatorname{Re}}
\renewcommand\Im{\operatorname{Im}}
\DeclareMathOperator{\Ref}{\mathfrak{R}}
\DeclareMathOperator{\Imf}{\mathfrak{I}}
\DeclareMathOperator{\im}{i}
\DeclareMathOperator{\eu}{e}
\DeclareMathOperator{\tr}{tr}
\DeclareMathOperator{\expm}{expm}
\DeclareMathOperator{\largestcoefficient}{largest\ coefficient}
\DeclareMathOperator{\sign}{sign}


\allowdisplaybreaks 

\makeatother

\providecommand{\examplename}{Example}
\providecommand{\lemmaname}{Lemma}
\providecommand{\remarkname}{Remark}
\providecommand{\theoremname}{Theorem}

\begin{document}

\title{Fast Numerical Nonlinear Fourier Transforms}

\author{Sander Wahls, \IEEEmembership{Member, IEEE,} and H. Vincent Poor,
\IEEEmembership{Fellow, IEEE}\thanks{S. Wahls is with the Delft Center for Systems and Control, Delft University
of Technology, Mekelweg 2, 2628 CD Delft, The Netherlands. Email:
\protect\href{http://s.wahls@tudelft.nl}{s.wahls@tudelft.nl}.

H. V. Poor is with the Department of Electrical Engineering, Princeton
University, Olden Street, Princeton, NJ 08544, USA. Email: \protect\href{http://poor@princeton.edu}{poor@princeton.edu}.}\thanks{This work was supported in part by the German Research Foundation
(DFG) under Grant WA 3139/1-1, and in part by the U. S. National Science
Foundation under Grant CCF-1420575.}}
\maketitle
\begin{abstract}
The nonlinear Fourier transform, which is also known as the forward
scattering transform, decomposes a periodic signal into nonlinearly
interacting waves. In contrast to the common Fourier transform, these
waves no longer have to be sinusoidal. Physically relevant waveforms
are often available for the analysis instead. The details of the transform
depend on the waveforms underlying the analysis, which in turn are
specified through the implicit assumption that the signal is governed
by a certain evolution equation. For example, water waves generated
by the Korteweg-de Vries equation can be expressed in terms of cnoidal
waves. Light waves in optical fiber governed by the nonlinear Schr\"odinger
equation (NSE) are another example. Nonlinear analogs of classic problems
such as spectral analysis and filtering arise in many applications,
with information transmission in optical fiber, as proposed by Yousefi
and Kschischang, being a very recent one. The nonlinear Fourier transform
is eminently suited to address them -- at least from a theoretical
point of view. Although numerical algorithms are available for computing
the transform, a ``fast'' nonlinear Fourier transform that is similarly
effective as the fast Fourier transform is for computing the common
Fourier transform has not been available so far. The goal of this
paper is to address this problem. Two fast numerical methods for computing
the nonlinear Fourier transform with respect to the NSE are presented.
The first method achieves a runtime of $O(D^{2})$ floating point
operations, where $D$ is the number of sample points. The second
method applies only to the case where the NSE is defocusing, but it
achieves an $O(D\log^{2}D)$ runtime. Extensions of the results to
other evolution equations are discussed as well. 
\end{abstract}


\section{Introduction}

Consider a smooth signal $q:\mathbb{R}\times\mathbb{R}_{\ge0}\to\mathbb{C}$
governed by the \emph{nonlinear Schr\"odinger equation (NSE)}\footnote{\label{fn:Second-form-of-the-NSE}Sometimes, the NSE (\ref{eq:NSE})
is given in the form $\im\partial_{t}u=\partial_{xx}u+2\kappa\vert u\vert^{2}u$.
This form is equivalent to (\ref{eq:NSE}) whenever $u=\bar{q}$.
Furthermore, the roles of the arguments $x$ and $t$ are sometimes
interchanged, e.g., when the NSE describes optical fiber. The spatial
variable is then commonly denoted by $z$ instead of $x$.} 
\begin{equation}
\im\partial_{t}q+\partial_{xx}q+2\kappa\vert q\vert^{2}q=0\label{eq:NSE}
\end{equation}
subject to a periodic boundary condition
\begin{equation}
q(x,t)\equiv q(x+\ell,t),\quad\ell>0.\label{eq:NSE-boundary-cond}
\end{equation}
The constant $\kappa\in\{\pm1\}$ determines whether the NSE is called
\emph{focusing} ($+$) or \emph{defocusing} ($-$). The NSE describes
several physically relevant phenomena. The complex envelope of a waveform
in an optical fiber with perfect loss compensation evolves according
to the NSE \cite[Ch. 6.1]{Hasegawa2003}. The focusing case corresponds
to a fiber with anomalous dispersion and admits \emph{bright solitons}
(``particle-like waves'') as solutions \cite[Ch. 5.1]{Hasegawa2003}.
Bright solitons are localized waves that remain unchanged after interactions
with other bright solitons. They are often employed to encode information
in optical communications \cite[Ch. 4]{Hasegawa2003}. The defocusing
case, which describes a fiber with normal dispersion, cannot be solved
by bright solitons. However, it admits solutions in the form of ``moving
holes'' in an otherwise constant signal, so-called \emph{dark solitons}
\cite[Ch. 5.4]{Hasegawa2003}. The use of dark solitons for optical
communications has been investigated much less than for bright solitons,
but the possibility of using them for optical communications has been
demonstrated experimentally \cite{Nakazawa1995}, \cite[p. 153ff]{Kivshar1998}.
The NSE also provides a model for the evolution of deep water waves
\cite{Craig1992}.

It was a pleasant surprise when Zakharov and Shabat \cite{Zakharov1972}
showed that the NSE (for non-periodic signals that decay sufficiently
rapidly as $|x|\to\infty$) can be solved in closed form using what
is known as the inverse scattering method. This method was initially
developed by Gardner et al. \cite{Gardner1967} for the solution of
the Korteweg-de Vries equation. Not much later, Ablowitz et al. \cite{Ablowitz1974}
extended the inverse scattering method to a wide range of evolution
equations. The method is usually called the inverse scattering method
because the main tools used in its derivation have their roots in
physics, where they are used to analyze how particles behave based
on their interactions with a scatterer. However, it can also be interpreted
as a generalization of the Fourier method for the solution of linear
evolution equations \cite{Ablowitz1974}. The direct time-evolution
of signals governed by such equations can be complicated, but the
time-evolution of their Fourier transforms often is simple. The inverse
scattering method exploits the same principle:
\[
\begin{array}{ccc}
 & \text{Nonlinear evol. eq.}\\
q(x,t_{0}) & \dashrightarrow & q(x,t_{1})\\
| &  & \uparrow\\
\text{Forward scattering} &  & \text{Inverse scattering}\\
\text{transform} &  & \text{transform}\\
\downarrow &  & |\\
\text{Scattering data} & \longrightarrow & \text{Scattering data}\\
\text{of }q(x,t_{0}) & \text{Simple(r)} & \text{of }q(x,t_{1})
\end{array}
\]
The forward scattering transform, which represents the signal in an
equivalent form called scattering data, can be seen as a generalization
of the Fourier transform because the scattering data essentially reduces
to the Fourier transform of the signal whenever the amplitude of the
signal, and hence the nonlinear term in the NSE, becomes small. Therefore
it is also known as the \emph{nonlinear Fourier transform (NFT)} in
the literature. 

Our interest in the NFT stems from three recent papers of Yousefi
and Kschischang on optical fiber communications \cite{Yousefi2012a,Yousefi2012b,Yousefi2013}.
The optical fiber channel suffers from several nonlinearities, most
of which are captured by the NSE. Current optical communication systems
treat the interference between multiple users that is caused by the
nonlinearities in a fiber as random noise. Motivated by the fact that
the data rates in current optical communication systems are close
to the capacity of that approach \cite{Essiambre2010,Winzer2012a},
Kschischang and Yousefi proposed a new method\footnote{Hasegawa \cite{Hasegawa1993} had proposed to embed data exlusively
in the solitonic part of the nonlinear Fourier spectrum already in
1993, but his proposal did not receive much attention. Recently, Prilepsky
et al. \cite{Prilepsky2013} proposed a system that utilizes exclusively
the non-solitonic part of the spectrum.} called \emph{nonlinear frequency division multiplexing (NFDM) }\cite{Yousefi2012a}\emph{.}
NFDM can be interpreted as a nonlinear variant of common \emph{orthogonal
frequency-division multiplexing (OFDM)} \cite{Yousefi2013}. The basic
idea is to generate the information-bearing signal in the scattering
domain in order to exploit the fact that the spatial evolution of
the scattering data is simpler than that of the original signal in
the time domain. One particular advantage of NFDM is that multi-user
interference can be avoided. At the receiver, the NFT is used to recover
the information. Yousefi and Kschischang investigated several numerical
methods in order to realize the NFT \cite{Yousefi2012b}, but found
that a computationally efficient fast NFT is still lacking \cite[Sec. VII]{Yousefi2012b}.
The goal of this paper is to address this problem. 

The authors have recently established the first fast nonlinear Fourier
transform for non-periodic signals governed by the focusing NSE that
decay sufficiently rapidly as $|x|\to\infty$ \cite{Wahls2013b}.
The same boundary conditions have been used by Yousefi and Kschischang,
but periodic boundary condition actually seem to be more appropriate
because one may use a cyclic prefix \cite[p. 156]{Yousefi2013c}.
In this paper, fast NFTs for signals governed by the periodic NSE
will be developed. While NFTs can be derived for signals governed
by many other different evolution equations as well \cite{Ablowitz1974},
we will mainly restrain ourselves to the NSE for the sake of clarity.
(Extensions to non-periodic signals and other evolution equations
will also be discussed, but only briefly.) Finally, let us note that
fast NFTs are of interest also in other areas. For example, they have
been used to analyze and filter water waves \cite{Osborne1993,Christov2009,Osborne2010,Bruehl2012}.
Measurements of oceanic data often contain up to $10.000$ data points
\cite[p. 95]{Osborne1993}. The high complexity of the nonlinear Fourier
transform has also been bemoaned in the analysis of plasma waves \cite[Sec. 8]{DudokdeWit1996}. 

The contributions of this paper are as follows. In Secs. \ref{sec:Finite-Band-Solutions}
and \ref{sec:The-NFT}, finite band solutions to the NSE are introduced
and a comprehensive survey of the relevant theoretical results on
the nonlinear Fourier transform, which is a method for extracting
the parameters of a finite band solution from measured data, is given.
The so-called monodromy matrix, which is an analytic matrix-valued
function, turns out to be the main tool in the derivation of the nonlinear
Fourier transform. The fast algorithms in this paper will require
a rational approximation of the monodromy matrix. The authors have
recently proposed a framework for obtaining rational approximations
of the analog of the monodromy matrix in the case of rapidly decaying
signals \cite{Wahls2013b}. In Sec. \ref{sec:Rational-Approximations-of-M},
this framework is carried over to the periodic case and extended by
the introduction of some new coordinate transforms. Then, in Sec.
\ref{sec:FNSFT-Eigen}, the fast algorithms presented in \cite{Wahls2013b}
are carried over to the periodic case and compared to other current
numerical approaches to find the nonlinear Fourier transform. In Sec.
\ref{sec:FNFT-Sampling}, a new alternative fast algorithm for the
defocusing NSE is introduced and compared. Some numerical examples
are presented in Sec. \ref{sec:Numerical-Examples}. Extensions of
our results to non-periodic signals and signals governed by other
evolution equations are discussed in the Secs. \ref{sec:Rapidly-Vanishing-Signals}
and \ref{sec:Other-Integrable-Evolution}, respectively. Finally,
the paper is concluded in Sec. \ref{sec:Conclusions}.

\subsection*{Notation}

Real numbers: $\mathbb{R}$; $\mathbb{R}_{\ge0}:=\{x\in\mathbb{R}:x\ge0\}$;
Complex numbers: $\mathbb{C}$; Integers: $\mathbb{Z}$; $\sqrt{\cdot}$:
Canonical square root (i.e., positive signs are preferred); $\im:=\sqrt{-1}$;
Euler's number: $\eu$; Real part: $\Re(\cdot)$; Imaginary part:
$\Im(\cdot)$; Complex conjugate: $\bar{(\cdot)}$; Natural logarithm:
$\ln(\cdot)$; Floor function: $\left\lfloor \cdot\right\rfloor $;
Absolute value: $\vert\cdot\vert$; Adjoint matrix: $(\cdot)^{*}$;
Matrix trace: $\tr\cdot$; Matrix exponential: $\expm(\cdot)$; Matrix
product: $\prod_{k=1}^{K}\mathbf{A}_{k}:=\mathbf{A}_{K}\mathbf{A}_{K-1}\times\dots\times\mathbf{A}_{1}$;
Matrix element in the $i$th column and $j$th row: $[\cdot]_{i,j}$;
Derivative w.r.t. a variable $u$: $\partial_{u}$; $\partial_{uv}:=\partial_{u}\partial_{v}$;
Equal for all arguments: $\equiv$; Absolute value of the largest
coefficient of a $\mathbb{C}^{m\times n}$-valued polynomial $\mathbf{p}(z)=\sum_{k=0}^{K}\mathbf{p}_{k}z^{k}$:
$|\largestcoefficient(\mathbf{p})|=\max\{|\left[\mathbf{p}_{k}\right]_{i,j}|:k=1,\dots,K,\,i=1,\dots,m,\,j=1,\dots,n\}$;
Degree of $\mathbf{p}$: $\deg(\mathbf{p})=\max\{k=1,\dots,K:\mathbf{p}_{k}\ne\mathbf{0}\}$

\section{Finite Band Solutions\label{sec:Finite-Band-Solutions}}

In this section, the theory of so-called  \emph{finite-band solutions}
to the NSE (\ref{eq:NSE}) will be reviewed \cite{Matveev2008}. Finite-band
solutions have explicit descriptions which rely only on a finite number
of parameters. These parameters will constitute the \emph{scattering
data}, and finding those parameters from an initial condition $q(x,t_{0})$
will constitute the non-linear Fourier transform as discussed in the
following section. Finite-band solutions can be loosely thought of
as the nonlinear analog to a conventional Fourier series expansion
with only finitely many non-zero terms. By restricting the exposition
to finite-band solutions, many questions regarding convergence can
be avoided. We remark that this means no significant loss of generality
since any sufficiently smooth periodic solution to the NSE can be
approximated arbitrarily well on any fixed finite time interval by
a periodic finite-band solution \cite{Grinevich2001}. In the literature,
two different types of finite band solutions can be found. Since each
type has its own advantages and disadvantages, both of them will be
reviewed in the following.

\subsection{The Finite-Band Solutions of Kotlyarov and Its\label{sub:Kotlyarov-finite-band}}

A solution $q(x,t)$ to the NSE (\ref{eq:NSE}) is called a \emph{finite-band
solution} in the sense of Kotlyarov and Its \cite{Kotlyarov1976,Its1976}
(also see \cite{Tracy1984,Tracy1984a}) if there exist finitely many
of the following parameters:
\begin{align}
\textit{Main spectrum:}\quad & \lambda_{1},\dots,\lambda_{2N}\in\mathbb{C},\label{eq:finite-gap-param-1}\\
\textit{Initial auxiliary spectrum:}\quad & \mu_{j}(x_{0},t_{0})\in\mathbb{C},\label{eq:finite-gap-param-2}\\
\textit{Initial amplitude\footnotemark:}\quad & q(x_{0},t_{0})\in\mathbb{R}_{\ge0},\label{eq:finite-gap-param-3}\\
\textit{Riemann sheet indices:}\quad & \sigma_{j}(x_{0},t_{0})\in\{\pm1\},\label{eq:finite-gap-param-5}\\
 & \quad j=1,\dots,N-1;\nonumber 
\end{align}
such that $q(x,t)$ is generated through the following system of coupled
partial differential equations:\footnotetext{\label{fn:phase-factor-NSE}The
restriction of $q(x_{0},t_{0})$ to be non-negative incurs no loss
of generality because if $q$ solves the NSE (\ref{eq:NSE}) then
so does $\eu^{\im\varphi}q$ for any $\varphi\in[0,2\pi)$.} \footnote{For $N=1$, empty sums in (\ref{eq:ODE-1})--(\ref{eq:ODE-3}) imply
zeros and empty products ones.}
\begin{align}
\partial_{x}\ln q= & 2\im\left(\sum_{j=1}^{N-1}\mu_{j}-\frac{1}{2}\sum_{k=1}^{2N}\lambda_{k}\right),\label{eq:ODE-1}\\
\partial_{t}\ln q= & 2\im\left(\sum_{{j,k=1\atop j>k}}^{2N}\lambda_{j}\lambda_{k}-\frac{3}{4}\left(\sum_{k=1}^{2N}\lambda_{k}\right)^{2}\right)\nonumber \\
 & +4\im\left(\frac{1}{2}\left(\sum_{k=1}^{2N}\lambda_{k}\right)\left(\sum_{j=1}^{N-1}\mu_{j}\right)-\sum_{{j,k=1\atop j>k}}^{N-1}\mu_{j}\mu_{k}\right)\label{eq:ODE-1a}\\
\partial_{x}\mu_{j}= & \frac{-2\im\sigma_{j}\sqrt{\prod_{k=1}^{2N}(\mu_{j}-\lambda_{k})}}{\prod_{{m=1\atop m\ne j}}^{N-1}(\mu_{j}-\mu_{m})},\label{eq:ODE-2}\\
\partial_{t}\mu_{j}= & -2\left(\sum_{{m=1\atop m\ne j}}^{N-1}\mu_{m}-\frac{1}{2}\sum_{k=1}^{2N}\lambda_{k}\right)\partial_{x}\mu_{j}.\label{eq:ODE-3}
\end{align}
The generated functions $\mu_{j}(x,t)$ are known as the \emph{auxiliary
spectrum}. The associated \emph{Riemann sheet indices} $\sigma_{j}(x,t)$
stay constant most of the time. Changes of sign occur if a $\mu_{j}(x,t)$
reaches one of the $\lambda_{k}$.\footnote{Technically, the $\mu_{j}$ evolve on a two-sheeted Riemann surface
specified by the $\lambda_{k}$ \cite[Apdx.]{Tracy1988}. The sheet
index indicates the current sheet.} These changes of sign result in a wave-like motion of the $\mu_{j}(x,t)$
that will oscillate between a pair of main spectral points $\lambda_{k}$
and $\lambda_{l}$, $k\ne l$. The $\mu_{j}(x,t)$ are also known
as \emph{hyperelliptic modes} in the literature \cite{Osborne2010}.
It may happen that a hyperelliptic mode is constant because it is
trapped between two repeating points in the main spectrum: $\mu_{j}(x,t)\equiv\lambda_{k}\equiv\lambda_{l}$
for $k\ne l$. The contributions of $\mu_{j}$, $\lambda_{k}$ and
$\lambda_{l}$ to (\ref{eq:ODE-1})--(\ref{eq:ODE-3}) cancel each
other out in that case, which implies these parameters do not contribute
to the shape of the function $q(x,t)$. Such parameters are called
\emph{degenerate}.  
\begin{rem}
[An assumption] \label{rem:assumption-main-spectrum}The representation
(\ref{eq:ODE-1})--(\ref{eq:ODE-3}) is not unique because it is possible
to add degenerate parameters without changing the solution. In order
to simplify the later exposition, we shall therefore assume (like,
e.g, in \cite{forest1986geometry}) that no point in the main spectrum
appears more than once:
\begin{equation}
k\ne l\Longrightarrow\lambda_{k}\ne\lambda_{l}.\label{eq:assumption-separate-main-spectrum}
\end{equation}
This condition in particular precludes the existence of degenerate
points in the main spectrum.
\end{rem}
At this point, it is interesting to note that $q(x,t)$ generated
through (\ref{eq:ODE-1})--(\ref{eq:ODE-3}) does not depend on whether
the NSE is defocusing or focusing (i.e., $\kappa=-1$ or $+1$). This
suggests that not every choice of the parameters (\ref{eq:finite-gap-param-1})--(\ref{eq:finite-gap-param-3})
will lead to a solution of the NSE, and this is indeed the case. The
following theorem can be used to check whether a given set of parameters
results in a (probably non-periodic) solution to the NSE.
\begin{thm}
[\cite{Tracy1984}; \cite{Tracy1984a}, Thm. 2.1]\label{thm:squared-eigenfunctions}The
function $q$ generated by the Eqs. (\ref{eq:ODE-1})--(\ref{eq:ODE-3})
solves the NSE (\ref{eq:NSE}), not necessarily subject to the boundary
condition (\ref{eq:NSE-boundary-cond}), if and only if the functions
\begin{align}
g_{z}(x,t):= & \im q(x,t)\prod_{j=1}^{N-1}(z-\mu_{j}(x,t)),\label{eq:squared-eigenfunction-g}\\
h_{z}(x,t):= & \im\kappa\bar{q}(x,t)\prod_{j=1}^{N-1}(z-\bar{\mu}_{j}(x,t)),\label{eq:squared-eigenfunction-h}\\
P(z):= & \prod_{k=1}^{2N}(z-\lambda_{k})\label{eq:squared-eigenfunction-P}
\end{align}
are such that the ``function-valued function'' 
\begin{equation}
z\mapsto f_{z}(x,t):=\sqrt{P(z)+g_{z}(x,t)h_{z}(x,t)}\label{eq:squared-eigenfunction-f}
\end{equation}
is a polynomial of finite degree. 
\end{thm}
The functions (\ref{eq:squared-eigenfunction-g}), (\ref{eq:squared-eigenfunction-h})
and (\ref{eq:squared-eigenfunction-f}) are known as the \emph{squared
eigenfunctions} in the literature \cite{forest1986geometry,Tracy1988}.
They play a fundamental role in the analysis of finite band signals
because their spatial and temporal evolution is remarkably simple.\footnote{\label{fn:evolution-sq-eigenfcts}There exist matrices $\boldsymbol{\Theta}_{z}(x,t)$
and $\boldsymbol{\Psi}_{z}(x,t)$, independent of $\boldsymbol{\theta}_{z}^{T}:=\left[\begin{array}{ccc}
f_{z} & g_{z} & h_{z}\end{array}\right]$, such that $\partial_{x}\boldsymbol{\theta}_{z}=\boldsymbol{\Theta}_{z}\boldsymbol{\theta}_{z}$
and $\partial_{t}\boldsymbol{\theta}_{z}=\boldsymbol{\Psi}_{z}\boldsymbol{\theta}_{z}$
\cite{Kotlyarov1976}, \cite[Eqs. 9+12]{Tracy1984}. An interesting
consequence is that a finite band signal $q(x,t)$ satisfies the NSE
(\ref{eq:NSE}) if and only $\partial_{xt}\boldsymbol{\theta}_{z}=\partial_{tx}\boldsymbol{\theta}_{z}$
\cite{forest1986geometry}.} Theorem \ref{thm:squared-eigenfunctions} can also serve as a starting
point for generating solutions to the NSE. This will be discussed
later in Sec. \ref{sub:Construction-of-Finite-Band-Solutions}.

Let us now illustrate the concepts introduced in this section so far
with a simple example.
\begin{example}
[Periodic one-band solution, $N=1$] \label{exa:Periodic-one-band-solution}
Theorem \ref{thm:squared-eigenfunctions} shows that $q(x,t)$ will
solve the NSE if and only if $f_{z}^{2}=P(z)+g_{z}h_{z}$ for some
unknown polynomial $z\mapsto f_{z}=\beta z-\gamma$: 
\begin{equation}
(\beta z-\gamma)^{2}=(z-\lambda_{1})(z-\lambda_{2})-\kappa|q|^{2}.\label{eq:f^2-eq-P-gh}
\end{equation}
By comparing the coefficients of the polynomials (with respect to
$z$) on both sides of (\ref{eq:f^2-eq-P-gh}), one finds $\beta=\pm1$
and $\gamma=\beta(\lambda_{1}+\lambda_{2})/2$. For both choices of
$\gamma$, a comparison of the constant terms in (\ref{eq:f^2-eq-P-gh})
results in the condition 
\begin{equation}
\left(\frac{\lambda_{1}-\lambda_{2}}{2}\right)^{2}=-\kappa|q|^{2}.\label{eq:example-cond-|q|}
\end{equation}

Since we are in the one-band case, this condition can also be obtained
directly. Solving (\ref{eq:ODE-1}) and (\ref{eq:ODE-1a}) for $N=1$
leads to 
\[
q(x,t)=q(x_{0},t_{0})\eu^{-\im(\lambda_{1}+\lambda_{2})(x-x_{0})}\eu^{2\im\left(\lambda_{1}\lambda_{2}-\frac{3}{4}(\lambda_{1}+\lambda_{2})^{2}\right)(t-t_{0})}.
\]
Direct substitution of this function into (\ref{eq:NSE}) shows again
that it solves the NSE if and only if (\ref{eq:example-cond-|q|})
is satisfied. Also note that the function is periodic in $x$ with
period $\ell$ {[}see (\ref{eq:NSE-boundary-cond}){]} if and only
if $\eu^{-\im(\lambda_{1}+\lambda_{2})\ell}=1$, or, equivalently,
$\frac{\ell}{2\pi}(\lambda_{1}+\lambda_{2})\in\mathbb{Z}$.
\end{example}
Finally, we note that finite-band solutions have a closed-form expression
which closely resembles the expression which arises when \emph{Hirota's
method} is used to solve the NSE with vanishing boundary conditions
\cite[Sec. III.B]{Yousefi2013}.

\begin{rem}
[Closed-form solution] \label{rem:closed-form-solution} Finite-band
solutions of the form (\ref{eq:ODE-1})--(\ref{eq:ODE-3}) can be
given in closed form \cite{Its1976}. The \emph{Riemann theta function}
with respect to a $d\times d$ matrix $\boldsymbol{\tau}$ is given
by 
\[
\Theta_{\boldsymbol{\tau}}(\mathbf{z}):=\sum_{\mathbf{m}\in\mathbb{Z}^{d}}\eu^{2\pi\im\mathbf{m}^{T}\mathbf{z}+\pi\im\mathbf{m}^{T}\boldsymbol{\tau}\mathbf{m}},\quad\mathbf{z}\in\mathbb{C}^{d}.
\]
The series converges absolutely for all $\mathbf{z}$ whenever $\boldsymbol{\tau}$
is symmetric with positive definite imaginary part. With a suitably
chosen $\boldsymbol{\tau}$, vectors $\mathbf{k}$, $\boldsymbol{\omega}$,
and $\boldsymbol{\delta}^{\pm}$, and scalars $k_{0}$ and $\omega_{0}$,
any finite band solution can be written as \cite[Eq. (A31)]{Tracy1988}
\begin{equation}
q(x,t)=q(x_{0},t_{0})\eu^{\im k_{0}x-\im\omega_{0}t}\frac{\Theta_{\boldsymbol{\tau}}(\frac{\pi}{2}(\mathbf{k}x+\boldsymbol{\omega}t+\boldsymbol{\delta}^{-}))}{\Theta_{\boldsymbol{\tau}}(\frac{\pi}{2}(\mathbf{k}x+\boldsymbol{\omega}t+\boldsymbol{\delta}^{+}))}.\label{eq:q(x,t)-via-theta}
\end{equation}
The parameters (\ref{eq:finite-gap-param-1})--(\ref{eq:finite-gap-param-5})
are the starting point if one wants to compute this representation
\cite[Apdx.]{Tracy1988} (also see \cite{Osborne2010,Tracy1984a}.)
A large family of parameters such that (\ref{eq:q(x,t)-via-theta})
leads to periodic solutions of the NSE (\ref{eq:NSE}) has been given
in \cite{Its1988}. An explicit parametrization of all parameters
that result in a periodic solution is given in \cite[Ch. 3]{Lee1986}.
A necessary and sufficient condition for periodic solutions with respect
to a discrete version of the NSE can be found in \cite[Thm. 5.2]{Miller1995}.
It should be mentioned that the naive evaluation of (\ref{eq:q(x,t)-via-theta})
becomes infeasible for larger $d$ due to the curse of dimensionality.
Numerical methods for the efficient evaluation of (certain) theta
functions have been discussed in \cite{Osborne2010,Deconinck2003,Osborne2010b}.
\end{rem}

\subsection{The Finite Band Solutions of Ma and Ablowitz\label{sub:Ma-finite-band}}

Finite band solutions as defined in Ma and Ablowitz \cite{Ma1981}
(also see \cite{Ma1977}) are specified through the following parameters:
\begin{align}
\textit{Main spectrum:}\quad & \lambda_{1},\dots,\lambda_{2N}\in\mathbb{C},\label{eq:finite-gap-main-spec-MaAbl}\\
\textit{Initial auxiliary spectra:}\quad & \varrho_{j}(x_{0},t_{0})\in\mathbb{C},\label{eq:finite-gap-aux-spec1-MaAbl}\\
 & \xi_{j}(x_{0},t_{0})\in\mathbb{C},\label{eq:finite-gap-aux-spec2-MaAbl}\\
\textit{Riemann sheet indices:}\quad & \nu_{j}(x_{0},t_{0})\in\{\pm1\},\label{eq:finite-gap-Riemann1-MaAbl}\\
 & \vartheta_{j}(x_{0},t_{0})\in\{\pm1\},\label{eq:finite-gap-Riemann2-MaAbl}\\
 & \quad j=1,\dots,N.\nonumber 
\end{align}
The auxiliary spectra evolve according to the following system of
coupled differential equations: 
\begin{align}
\partial_{x}\varrho_{j}= & \frac{2\im\nu_{j}\sqrt{\prod_{k=1}^{2N}(\varrho_{j}-\lambda_{k})}}{\prod_{{i=1\atop i\ne j}}^{N}(\varrho_{j}-\varrho_{i})}\left(\sum_{{i=1\atop i\ne j}}^{N}\varrho_{i}-\frac{1}{2}\sum_{k=1}^{2N}\lambda_{k}\right),\label{eq:aux-spec-MaAblo1-evol}\\
\partial_{x}\xi_{j}= & \frac{2\im\vartheta_{j}\sqrt{\prod_{k=1}^{2N}(\xi_{j}-\lambda_{k})}}{\prod_{{i=1\atop i\ne j}}^{N}(\xi_{j}-\xi_{i})}\left(\sum_{{i=1\atop i\ne j}}^{N}\xi_{i}-\frac{1}{2}\sum_{k=1}^{2N}\lambda_{k}\right).\label{eq:aux-spec-MaAblo2-evol}
\end{align}
There are also differential equations that govern the temporal evolution
of the auxiliary spectra \cite[Eqs. (4.2)+(8.4)]{Ma1981}, but they
are quite complicated and will therefore not be given here. As before,
the auxiliary spectra change their sheet indices if and only if one
of them reaches a point in the main spectrum. The corresponding finite-band
signal is given by
\begin{equation}
q=\frac{\kappa+\im}{2}\sum_{k=1}^{2N}\lambda_{k}-\kappa\sqrt{-\kappa}\sum_{j=1}^{N}\varrho_{j}-\sqrt{\kappa}\sum_{j=1}^{N}\xi_{j}.\label{eq:finite-gap-signal-MaAblo}
\end{equation}

Again, not every choice of initial parameters will lead to a solution
of the NSE. In contrast to the finite-band solutions of Kotlyarov
and Its, no condition for that seems to be known. The reconstruction
formula (\ref{eq:finite-gap-signal-MaAblo}) is now simpler, but the
evolution of the auxiliary spectra (\ref{eq:finite-gap-aux-spec1-MaAbl}),
(\ref{eq:finite-gap-aux-spec2-MaAbl}) is more complicated. The main
spectrum coincides with that in Sec. \ref{sub:Kotlyarov-finite-band}.
\begin{example}
[One band; \cite{Ma1981}, Sec. 2.4] In the Ma-Ablowitz case, finding
the auxiliary spectra $\varrho_{j}$ and $\xi_{j}$ is complicated
even in the one-band case $N=1$. The general form of $q(x,t)$ for
$N=1$ again turns out to be \cite[p. 133ff]{Ma1981}
\[
q(x,t)=A\eu^{\kappa\im\frac{c_{0}}{c_{1}}x}\eu^{-\im((\frac{c_{0}}{c_{1}})^{2}+2\kappa A^{2})t}.
\]
The free parameters $A,c_{0},c_{1}$ are related to the spectral parameters
as follows:
\begin{align*}
\lambda_{1}= & \frac{c_{0}}{2c_{1}}-\sqrt{-\kappa}A,\quad\lambda_{2}=\frac{c_{0}}{2c_{1}}+\sqrt{-\kappa}A,\\
\varrho_{1}(x,t)= & \frac{c_{0}}{2c_{1}}-\sqrt{-\kappa}A\sin\left(\frac{c_{0}}{c_{1}}x+\left(-\kappa(\frac{c_{0}}{c_{1}})^{2}+2A^{2}\right)t\right),\\
\xi_{1}(x,t)= & \frac{c_{0}}{2c_{1}}-\sqrt{-\kappa}A\cos\left(\frac{c_{0}}{c_{1}}x+\left(-\kappa(\frac{c_{0}}{c_{1}})^{2}+2A^{2}\right)t\right).
\end{align*}

\end{example}
\begin{lyxlist}{00.00.0000}
\item [{}]~
\end{lyxlist}

\subsection{Construction of Finite-Band Solutions for Information Transmission
in Optical Fiber\label{sub:Construction-of-Finite-Band-Solutions}}

The efficient construction of signals with prescribed scattering data
is fundamental in optical communication systems based on nonlinear
Fourier transforms, where this problem corresponds to generating the
input to the fiber on the transmitter side. The first papers addressing
this problem have been published only very recently \cite{Yousefi2013,Hari2014,Wahls2014a,Turitsyna2013,Prilepsky2013,Prilepsky2014a,Le2014,Zhang2014},
all for the NSE with vanishing boundary conditions. We remark that
in all these works, not all available degrees of freedom are exploited
in order to reduce the computational complexity of the problem. It
has also been observed that some degrees of freedom seem to be ill-suited
for information transmission due to sensitivity issues \cite[VIII]{Yousefi2012b}.
Another problem in these works arises due to the chosen boundary conditions:
it is difficult to control the temporal spread of the generated signals.
The only established method so far seems to be pruning of the signal
set \cite[Sec. V.C]{Yousefi2013}, \cite[Sec. III]{Hari2014}, which
is only feasible if the number of degrees of freedom is small. 

It seems that the construction of periodic solutions to the NSE with
prescribed scattering data for optical communication has not yet been
discussed in the literature. We therefore now review quickly a few
potential starting points. The construction of periodic finite-band
signals appears at first more complicated than for vanishing boundary
conditions because the parameters (\ref{eq:finite-gap-param-1})--(\ref{eq:finite-gap-param-5})
are coupled through the condition in Theorem \ref{thm:squared-eigenfunctions}.
However, they offer the important advantage that their temporal support
is fixed. Theorem \ref{thm:squared-eigenfunctions} can serve as a
starting point for the construction of parameters (\ref{eq:finite-gap-param-1})--(\ref{eq:finite-gap-param-5})
such that the function $q(x,t)$ generated by (\ref{eq:ODE-1})--(\ref{eq:ODE-3})
actually solves the NSE (\ref{eq:NSE}) \cite{Tracy1984,Tracy1984a},
but it seems difficult to enforce the periodic boundary condition
(\ref{eq:NSE-boundary-cond}), especially if a specific period is
desired. However, several methods based on the \emph{Darboux-B\"acklund
transforms} (which have also been employed for vanishing boundary
conditions \cite{Yousefi2013}) are available for the construction
of periodic finite-gap solutions to the NSE with pre-specified (complex)
main spectrum: see \cite{Steudel1986}, \cite[Ch. 3.1]{Schober1991}
and \cite[Sec. 4.2]{Li1994}, and \cite[Thm. 6.15]{Cascaval2004}.
The theta function representation discussed in Remark \ref{rem:closed-form-solution}
offers another potential way to generate finite-band solutions. The
advantage of these approaches it that the period of the constructed
solution, and therefore its temporal spread, can be controlled.

\section{The Nonlinear Fourier Transform\label{sec:The-NFT}}

The NFT of a \emph{periodic} finite-band solution $q$, taken at a
reference time $t_{0}$ and with respect to a reference point $x_{0}$,
is the mapping from $q(\cdot,t_{0})$ and $x_{0}$ to the scattering
data, which is given by either (\ref{eq:finite-gap-param-1})--(\ref{eq:finite-gap-param-5}),
or (\ref{eq:finite-gap-main-spec-MaAbl})--(\ref{eq:finite-gap-Riemann2-MaAbl}).
In this section, the computation of the scattering data is reviewed
as a preparation for the derivation of the numerical algorithms. First,
the main points of the Lax pair formalism are explained in order to
motivate the spectral analysis of the differential operator
\begin{equation}
\mathbf{L}_{t_{0}}=\im\left[\begin{array}{cc}
\frac{d}{dx} & q(\cdot,t_{0})\\
\kappa\bar{q}(\cdot,t_{0}) & -\frac{d}{dx}
\end{array}\right].\label{eq:L}
\end{equation}
Second, the spectral theory for the operator $\mathbf{L}_{t_{0}}$
will be reviewed. In particular, the monodromy matrix and the Floquet
discriminant will be introduced. Finally, the relation between the
spectrum of $\mathbf{L}_{t_{0}}$ and the scattering data will be
established.
\begin{rem}
In the literature, $\mathbf{L}_{t_{0}}$ is sometimes replaced with
\[
\left[\begin{array}{cc}
1\\
 & \zeta
\end{array}\right]\mathbf{L}_{t_{0}}\left[\begin{array}{cc}
1\\
 & \zeta
\end{array}\right]^{-1}=\left[\begin{array}{cc}
\im\frac{d}{dx} & \frac{\im}{\zeta}q(\cdot,t_{0})\\
\im\zeta\kappa\bar{q}(\cdot,t_{0}) & -\im\frac{d}{dx}
\end{array}\right].
\]
All these operators are similar. Their eigenvalues coincide.
\end{rem}

\subsection{The Lax Pair Formalism\label{sub:Lax-Pair-Formalism}}

The relation between the operator $\mathbf{L}_{t_{0}}$ and the NSE
(\ref{eq:NSE}) is as follows. One can find a second differential
operator $\mathbf{B}$, which also depends on $q$, such that the
condition
\begin{equation}
\partial_{t_{0}}\mathbf{L}_{t_{0}}=\mathbf{B}\mathbf{L}_{t_{0}}-\mathbf{L}_{t_{0}}\mathbf{B}\label{eq:Lax-condition}
\end{equation}
is equivalent to $q$ being a solution to the NSE \cite{Zakharov1972}.
The details of how to find a suitable operator $\mathbf{B}$ will
not be given here because this procedure is not important in this
paper. See \cite[Ch. 6.1]{Eckhaus1981} for an in-depth derivation.
Any two operators $\mathbf{L}_{t_{0}}$ and $\mathbf{B}$ that satisfy
the condition (\ref{eq:Lax-condition}) are said to form a \emph{Lax
pair}. The main point about Lax pairs is that the eigenvalues of $\mathbf{L}_{t_{0}}$
are independent of $t_{0}$ \cite{Lax1968}. Furthermore, the time-evolution
of any eigenfunction $\boldsymbol{\phi}_{t_{0}}(x)$ of $\mathbf{L}_{t_{0}}$
is simply $\partial_{t_{0}}\boldsymbol{\phi}_{t_{0}}=\mathbf{B}\boldsymbol{\phi}_{t_{0}}$.
This relation is the reason why the time-evolution of the scattering
data, which will be derived from the eigenstructure of $\mathbf{L}_{t_{0}}$,
is usually  simpler than that of the original signal.

\subsection{Monodromy Matrix and Floquet Discriminant}

The eigenproblem $\mathbf{L}_{t_{0}}\boldsymbol{\upsilon}=z\boldsymbol{\upsilon}$
can be rearranged to
\begin{equation}
\frac{d}{dx}\boldsymbol{\upsilon}=\left[\begin{array}{cc}
-\im z & -q(\cdot,t_{0})\\
\kappa\bar{q}(\cdot,t_{0}) & \im z
\end{array}\right]\boldsymbol{\upsilon}.\label{eq:Lv-lamv-as-DE}
\end{equation}
Equation (\ref{eq:Lv-lamv-as-DE}) has a unique non-trivial solution
for any initial condition of the form $\boldsymbol{\upsilon}(x_{0})=\boldsymbol{\upsilon}_{0}\ne\mathbf{0}$
\cite[Thm. 3.9]{Teschl2010}. However, although the coefficients in
(\ref{eq:Lv-lamv-as-DE}) are periodic, solutions to (\ref{eq:Lv-lamv-as-DE})
will not be periodic in general. The choice $q\equiv0$ and $z=\im$,
for example, leads to $\boldsymbol{\upsilon}(x)^{T}=\left[\begin{array}{cc}
\eu^{x} & \eu^{-x}\end{array}\right]$. In the following, only eigenvalues that admit bounded quasi-periodic
eigenfunctions (i.e., $\boldsymbol{\upsilon}(x+\ell)\equiv m\boldsymbol{\upsilon}(x)$
with $|m|=1$) will be of interest. The analysis of differential equations
with periodic coefficients is the subject of Floquet theory \cite[Ch. 3.6]{Teschl2010}.
Following \cite[Apdx.]{Tracy1988}, we now outline how Floquet theory
allows us to identify these eigenvalues in a way that is similar to
how eigenvalues with finite-energy eigenfunctions are found for vanishing
boundary conditions \cite{Yousefi2012a}. 

Let $\boldsymbol{\phi}_{x_{0},t_{0},z}$ and $\tilde{\boldsymbol{\phi}}_{x_{0},t_{0},z}$
denote the solutions of Eq. (\ref{eq:Lv-lamv-as-DE}) with respect
to the canonical initial conditions
\begin{equation}
\boldsymbol{\phi}_{x_{0},t_{0},z}(x_{0})=\left[\begin{array}{c}
1\\
0
\end{array}\right],\quad\tilde{\boldsymbol{\phi}}_{x_{0},t_{0},z}(x_{0})=\left[\begin{array}{c}
0\\
1
\end{array}\right]\label{eq:boundary-conditions-canonical-evs}
\end{equation}
and the argument $z$. The \emph{monodromy matrix} 
\begin{equation}
\mathbf{M}_{x_{0},t_{0}}(z):=\left[\begin{array}{cc}
\boldsymbol{\phi}_{x_{0},t_{0},z}(x_{0}+\ell) & \tilde{\boldsymbol{\phi}}_{x_{0},t_{0},z}(x_{0}+\ell)\end{array}\right]\label{eq:monodromy-matrix}
\end{equation}
captures the evolution of these two solutions over one period. It
can be thought of as the equivalent of the transfer matrix used with
vanishing boundary conditions \cite{Yousefi2012a}. The monodromy
matrix allows us to identify the $z$ that admit quasi-periodic eigenfunctions
with desired period transitions as follows.
\begin{lem}
[\cite{Tracy1988}, p. 831f.]\label{lem:monodromy-matrix-quasiperiodic-ev}Fix
two arbitrary complex constants $z,m\in\mathbb{C}$. Then, the eigenproblem
$\mathbf{L}_{t_{0}}\boldsymbol{\upsilon}=z\boldsymbol{\upsilon}$
admits a quasi-periodic eigenfunction $\boldsymbol{\upsilon}\ne\mathbf{0}$
in the sense that $\boldsymbol{\upsilon}(x+\ell)\equiv m\boldsymbol{\upsilon}(x)$
if and only if $\Delta(z):=\frac{1}{2}\tr\mathbf{M}_{x_{0},t_{0}}(z)$
satisfies
\begin{equation}
m^{2}-2m\Delta(z)+1=0.\label{eq:condition-for-quasiperiodic-ev}
\end{equation}

\end{lem}
The function $\Delta$ is known as \emph{Floquet discriminant} in
the literature. As the notation suggests, $\Delta$ indeed does not
depend on the reference points $x_{0}$ and $t_{0}$. 
\begin{lem}
[\cite{Ma1981}, Eqs. (3.8)+(8.3)]\label{lem:Floquet-discriminant-is-independent}
The Floquent discriminant $\Delta$ is independent of the reference
points $x_{0}$ and $t_{0}$.
\end{lem}
The determinant of the monodromy matrix is also invariant.
\begin{lem}
[\cite{Ma1981}, Eqs. (1.5)+(6.3c)]\label{lem:detM-is-one}The monodromy
matrix satisfies $\det\mathbf{M}_{x_{0},t_{0}}(z)\equiv1$ for all
$x_{0}$ and $t_{0}$.
\end{lem}
The monodromy matrix possesses some symmetries.
\begin{lem}
[\cite{Ma1981}, Secs. I.1+II.1]\label{lem:monodromy-symmetry-like-props}We
have $\left[\mathbf{M}_{x_{0},t_{0}}(z)\right]_{2,2}\equiv\left[\bar{\mathbf{M}}_{x_{0},t_{0}}(\bar{z})\right]_{1,1}$
and $\left[\mathbf{M}_{x_{0},t_{0}}(z)\right]_{2,1}\equiv-\kappa\left[\bar{\mathbf{M}}_{x_{0},t_{0}}(\bar{z})\right]_{1,2}$.
\end{lem}

\subsection{Connection between Kotlyarov-Its Finite-Band Solutions and the Monodromy
Matrix}

In this subsection, we assume that $q(x,t)$ is a periodic finite-band
solution in the sense of Kotlyarov and Its. We develop expressions
which allow us to compute the corresponding parameters (\ref{eq:finite-gap-param-1})--(\ref{eq:finite-gap-param-5})
with the help of the monodromy matrix.

\subsubsection{Squared Eigenfunctions in Terms of the Monodromy Matrix}

The following theorem will enable us to evaluate the squared eigenfunctions
given in Eqs. (\ref{eq:squared-eigenfunction-g}), (\ref{eq:squared-eigenfunction-h})
and (\ref{eq:squared-eigenfunction-f}) indirectly through evaluations
of the monodromy matrix up to an unknown but commonly shared non-zero
factor.
\begin{thm}
[\cite{Kotlyarov1976}; \cite{Tracy1984a}, Sec. 4.2, Apdx. I; \cite{Tracy1991},
Sec. 4.6]\label{thm:Squared-Eigfun-Monodromy-Mat} Fix any choice
of squared eigenfunctions (\ref{eq:squared-eigenfunction-g}), (\ref{eq:squared-eigenfunction-h})
and (\ref{eq:squared-eigenfunction-f}) such that the associated parameters
(\ref{eq:finite-gap-param-1})--(\ref{eq:finite-gap-param-5}) generate
$q(x,t)$. Then, there exists a function $C:\mathbb{C}\to\mathbb{C}$
such that
\begin{align*}
C(z)f_{z}(x_{0},t_{0})= & -\im(\Delta(z)-\left[\mathbf{M}_{x_{0},t_{0}}(z)\right]_{1,1}),\\
C(z)g_{z}(x_{0},t_{0})= & \left[\mathbf{M}_{x_{0},t_{0}}(z)\right]_{1,2},\\
C(z)h_{z}(x_{0},t_{0})= & \left[\mathbf{M}_{x_{0},t_{0}}(z)\right]_{2,1}.
\end{align*}
\end{thm}
\begin{IEEEproof}
We only sketch the main idea. One defines
\begin{align*}
\tilde{f}_{z}(x,t):= & -\frac{\im}{2}([\boldsymbol{\phi}_{x,t,z}]_{1}[\tilde{\boldsymbol{\phi}}_{x,t,z}]_{2}+[\boldsymbol{\phi}_{x,t,z}]_{2}[\tilde{\boldsymbol{\phi}}_{x,t,z}]_{1}),\\
\tilde{g}_{z}(x,t):= & [\boldsymbol{\phi}_{x,t,z}]_{1}[\tilde{\boldsymbol{\phi}}_{x,t,z}]_{1},\;\tilde{h}_{z}:=-\kappa[\boldsymbol{\phi}_{x,t,z}]_{2}[\tilde{\boldsymbol{\phi}}_{x,t,z}]_{2}.
\end{align*}
Then, one uses the fact that $\boldsymbol{\phi}_{x_{0},t_{0},z}$
and $\tilde{\boldsymbol{\phi}}_{x_{0},t_{0},z}$ are eigenfunctions
of $\mathbf{L}_{t_{0}}$ to show that the spatio-temporal evolution
of $\tilde{\boldsymbol{\theta}}_{z}^{T}:=\left[\begin{array}{ccc}
\tilde{f}_{z} & \tilde{g}_{z} & \tilde{h}_{z}\end{array}\right]$ is also governed by same differential equations that were mentioned
earlier for $\boldsymbol{\theta}_{z}$ in Footnote \ref{fn:evolution-sq-eigenfcts}.
Uniqueness arguments then show that $\boldsymbol{\theta}_{z}$ and
$\tilde{\boldsymbol{\theta}}_{z}$ differ only by a normalization
factor. Eventually, one connects $\tilde{f}_{z}$, $\tilde{g}_{z}$
and $\tilde{h}_{z}$ to the monodromy matrix using (\ref{eq:monodromy-matrix}).
\end{IEEEproof}

\subsubsection{Main Spectrum \label{sub:Main-Spectrum}}

Solving for $m=\pm1$ in Eq. (\ref{eq:condition-for-quasiperiodic-ev})
shows that the operator $\mathbf{L}_{t_{0}}$ admits an (anti-)periodic
eigenfunction for some $z$ if and only if the Floquet discriminant
satisfies $\Delta(z)\in\{\pm1\}$. Under some weak assumptions, Theorem
\ref{thm:squared-eigenfunctions} implies that the main spectrum of
a finite-band solution corresponds to the simple (anti-)periodic eigenvalues
of $\mathbf{L}_{t_{0}}$.

\begin{lem}
\label{lem:Main-spectrum-Via-Floquet-Discriminant} Assume that there
is a finite-band representation for $q(x,t)$ that satisfies (\ref{eq:assumption-separate-main-spectrum}),
and fix it. Furthermore, assume that the roots of $1-\Delta^{2}$
are at most double. The main spectrum (\ref{eq:finite-gap-param-1})
then corresponds exactly to the simple roots of $1-\Delta^{2}$: 
\begin{equation}
\{\lambda_{k}\}_{k=1}^{2N}=\left\{ \zeta\in\mathbb{C}:\Delta(\zeta)\in\{\pm1\},\frac{d\Delta}{dz}(\zeta)\ne0\right\} .\label{eq:main-spec-via-floquet}
\end{equation}
\end{lem}
\begin{IEEEproof}
The main spectrum corresponds to the simple roots of $P=f_{z}^{2}-g_{z}h_{z}$
{[}cf. (\ref{eq:assumption-separate-main-spectrum}), (\ref{eq:squared-eigenfunction-P})
and (\ref{eq:squared-eigenfunction-f}){]}. With the help of Lemma
\ref{lem:detM-is-one}, Theorem \ref{thm:Squared-Eigfun-Monodromy-Mat}
can be used to show that $C^{2}P=(1-\Delta^{2})$. Now, every simple
root of $1-\Delta^{2}$ must be a root of $P$ because the roots of
$C^{2}$ are at least double. On the other hand, if $P$ has a root
then $C^{2}$ cannot have a root because the roots of $1-\Delta^{2}$
are at most double. Thus, every simple root of $P$ is also a simple
root of $1-\Delta^{2}$.
\end{IEEEproof}
A natural question arising at this point is whether the non-simple
roots of $1-\Delta^{2}$ are of any importance. The answer is yes;
they are essential for analyzing the impact of perturbations.
\begin{rem}
[Non-simple roots] \label{rem:non-simple-roots} The non-simple roots\footnote{i.e., $\zeta\in\mathbb{C}$ such that $\Delta(\zeta)\in\{\pm1\}$
and $\frac{d\Delta}{dz}(\zeta)=0$.} of $1-\Delta^{2}$ can be interpreted as ``canonical'' degenerate
points in the main spectrum of a finite-band solution. Since these
roots are eigenvalues of the operator $\mathbf{L}_{t_{0}}$, perturbation
theory shows that double-roots will in general split up into two simple
roots, leading to new non-degenerate points in the main spectrum.
The corresponding hyperelliptic mode $\mu_{j}$ is no longer trapped.
Although the impact of small perturbations on the roots themselves
is small, they can nevertheless change the trajectories of the formerly
trapped hyperelliptic mode significantly \cite{Tracy1984a,Tracy1988,forest1986geometry,Lee1986}.
As the solution evolves, this can lead to instabilities known as \emph{rogue
(or freak) waves} \cite{Osborne2010b,Solli2007}. Whether or not a
double root can lead to an instability depends on its location in
the complex plane. We note that real double roots cannot cause instabilities
\cite[Thm. 5.3]{forest1986geometry}.
\end{rem}
Let us now illustrate matters with another example.
\begin{example}
[Plane wave; e.g. \cite{Tracy1988}, Sec. II.A] \label{exa:plane-wave}Consider
the following periodic solution to the focusing NSE: 
\[
q(x,t)=q_{0}\eu^{2\im q_{0}^{2}t},\quad q_{0}\ge0.
\]
This is a special case of Ex. \ref{exa:Periodic-one-band-solution}.
We therefore know that a one-band representation for $q(x,t)$ that
satisfies (\ref{eq:assumption-separate-main-spectrum}) exists, and
can use Lem. \ref{lem:Main-spectrum-Via-Floquet-Discriminant} to
analyze this particular representation. The Floquet discriminant is
$\Delta(z)=\cos(\ell\sqrt{z^{2}+q_{0}^{2}})$, which leads to infinitely
many roots for $\Delta(z)\pm1$: 
\[
\zeta_{n}^{\pm}:=\pm\sqrt{n^{2}\pi^{2}/\ell^{2}-q_{0}^{2}},\quad n\in\mathbb{N}.
\]
By taking the limit $\zeta\to\zeta_{n}^{\pm}$ with respect to 
\[
\frac{d\Delta}{dz}(\zeta)=-2\zeta\ell\frac{\sin\left(\ell\sqrt{\zeta^{2}+q_{0}^{2}}\right)}{\sqrt{\zeta^{2}+q_{0}^{2}}},
\]
one finds that only the roots at $\zeta_{0}^{\pm}=\pm\im q_{0}$ are
simple. Thus, the main spectrum is $\lambda_{1}=\zeta_{0}^{+}$ and
$\lambda_{2}=\zeta_{0}^{-}$. The non-simple roots correspond to the
$\zeta_{n}^{\pm}$ with $n\ge1$. They are imaginary if $n<q_{0}\ell/\pi$
and real otherwise. The effect of perturbations on the non-simple
roots is discussed e.g. in \cite{Tracy1988,forest1986geometry}.
\end{example}
In this example, it is interesting to note that the non-degenerate
main spectrum is symmetric with respect to the real axis, and that
the number of non-real double roots is finite. These two properties
can be generalized as follows. 
\begin{lem}
[\cite{forest1986geometry,Ma1977}]\label{lem:main-spectrum-symmetry}
If the main spectrum is given by (\ref{eq:main-spec-via-floquet}),
it must consist of complex conjugate pairs. Furthermore, $\{\lambda_{j}\}_{j=1}^{2N}\subset\mathbb{R}$
in the defocusing case $\kappa=-1$.\end{lem}
\begin{IEEEproof}
Lemma \ref{lem:monodromy-symmetry-like-props} implies $\bar{\Delta}(\bar{z})\equiv\Delta(z)$,
and thus (\ref{eq:main-spec-via-floquet}) is symmetric with respect
to the real axis. The $\lambda_{j}$ are real if $\kappa=-1$ because
$\mathbf{L}_{t_{0}}$ then is self-adjoint with respect to $\langle\boldsymbol{\phi},\tilde{\boldsymbol{\phi}}\rangle=\int_{x_{0}}^{x_{0}+\ell}\tilde{\boldsymbol{\phi}}(x)^{*}\boldsymbol{\phi}(x)dx$.
\end{IEEEproof}
{}
\begin{lem}
[\cite{Tracy1988,forest1986geometry}] \label{lem:finitely-many-imaginary-roots}
The functions $\Delta(z)\pm1$ have only finitely many non-simple
non-real roots. That is, $1-\Delta(\zeta)^{2}=\frac{d\Delta}{dz}(\zeta)=0$
for only finitely many complex points $\zeta\notin\mathbb{R}$.
\end{lem}
Lemma \ref{lem:finitely-many-imaginary-roots} is important because,
as mentioned in Remark \ref{rem:non-simple-roots}, only non-real
double roots can lead to instabilities \cite{Tracy1988,forest1986geometry}.
Thus, only finitely many degenerate points have to be taken into account
during a stability analysis.

\subsubsection{Auxiliary Spectrum }

Next, consider the $z$ for which the first element of the canonical
eigenfunction $\tilde{\boldsymbol{\phi}}_{x_{0},t_{0},z}$ vanishes
at $x_{0}$ and $x_{0}+\ell$: 
\[
\left[\tilde{\phi}_{x_{0},t_{0},z}(x_{0})\right]_{1}=\left[\tilde{\phi}_{x_{0},t_{0},z}(x_{0}+\ell)\right]_{1}=0.
\]
By definition of the monodromy matrix, this condition corresponds
to the upper right element of the monodromy matrix being zero. The
following lemma implies that these $z$ constitute the auxiliary spectrum
discussed in Sec. \ref{sub:Kotlyarov-finite-band} up to degenerate
parts which can be canceled as follows.
\begin{lem}
\label{lem:aux-spectrum-via-monodromy}Let $q(x_{0},t_{0})\ne0$,
and assume (\ref{eq:assumption-separate-main-spectrum}) and that
the roots of $1-\Delta^{2}$ are at most double. Furthermore, set
$n(\zeta):=1$ if $\zeta\in\mathbb{C}$ is a double root of $1-\Delta^{2}$
and $n(\zeta):=0$ otherwise. Then, the auxiliary spectrum is
\[
\{\mu_{j}(x_{0},t_{0})\}_{j=1}^{N-1}=\left\{ \zeta\in\mathbb{C}:\lim_{z\to\zeta}\frac{\left[\mathbf{M}_{x_{0},t_{0}}(z)\right]_{1,2}}{(z-\zeta)^{n(\zeta)}}=0\right\} .
\]
\end{lem}
\begin{IEEEproof}
In the proof of Lem. \ref{lem:Main-spectrum-Via-Floquet-Discriminant},
it was shown that $C^{2}=(1-\Delta^{2})/P$ and that the roots of
$P$ are exactly the simple roots of $1-\Delta^{2}$. Thus, the roots
of $C^{2}=(1-\Delta^{2})/P$ are exactly the remaining double roots
of $1-\Delta^{2}$. The claim follows from (\ref{eq:squared-eigenfunction-g})
because $Cg_{z}=\left[\mathbf{M}_{x_{0},t_{0}}(z)\right]_{1,2}$ by
Thm. \ref{thm:Squared-Eigfun-Monodromy-Mat}.
\end{IEEEproof}

\subsection{Connection between Ma-Ablowitz Finite-Band Solutions and the Monodromy
Matrix\label{sub:Ma-Ablowitz}}

In this subsection, the approach in Ma and Ablowitz \cite{Ma1981}
is reviewed and some expressions that will be convenient later are
derived. Ma and Ablowitz assume that a solution $q(x,t)$ to the periodic
NSE (\ref{eq:NSE})--(\ref{eq:NSE-boundary-cond}) is given such that
$1-\Delta^{2}$ has only finitely many simple roots. In the focusing
case, it is furthermore assumed that:\footnote{These assumptions are quite strong. The function in Example \ref{exa:plane-wave},
e.g., has non-real double roots and therefore  violates them.} \cite[p. 129f]{Ma1981}
\begin{enumerate}
\item all real roots of $1-\Delta^{2}$ are double,
\item all non-real roots of $1-\Delta^{2}$ are simple, 
\item all roots of the terms defining the auxiliary spectra {[}i.e., (\ref{eq:MaAblow-aux-spec-via-roots-1})
and (\ref{eq:MaAblow-aux-spec-via-roots-2}) below{]} are simple,
and
\item all real roots of the terms defining the auxiliary spectra coincide
with a root of $1-\Delta^{2}$.
\end{enumerate}
Under these conditions, Ma and Ablowitz prove that the following finite-band
solution coincides with $q(x,t)$ \cite{Ma1981}.

The (non-degenerate) main spectrum of the finite-band representation
of $q(x,t)$ has been defined as the simple roots of $1-\Delta^{2}$,
i.e., again via (\ref{eq:main-spec-via-floquet}) \cite[p. 116f+129]{Ma1981}.
The corresponding auxiliary spectra (\ref{eq:finite-gap-aux-spec1-MaAbl})--(\ref{eq:finite-gap-aux-spec2-MaAbl})
have been defined as the solutions (with respect to $\varrho$ and
$\xi$) of \cite[p. 116f+129]{Ma1981} 
\begin{align}
\Imf[\phi_{x_{0},t_{0},\varrho,1}(x_{0}+\ell)]-\im\sqrt{\kappa}\Imf[\phi_{x_{0},t_{0},\varrho,2}(x_{0}+\ell)] & =0,\label{eq:MaAblow-aux-spec-via-roots-1}\\
\Imf[\phi_{x_{0},t_{0},\xi,1}(x_{0}+\ell)]-\im\sqrt{\kappa}\Ref[\phi_{x_{0},t_{0},\xi,2}(x_{0}+\ell)] & =0,\label{eq:MaAblow-aux-spec-via-roots-2}
\end{align}
where $\Ref[\phi(z)]:=\frac{1}{2}(\phi(z)+\bar{\phi}(\bar{z}))$ and
$\Imf[\phi(z)]:=\frac{1}{2\im}(\phi(z)-\bar{\phi}(\bar{z}))$. These
conditions can be rewritten using the functions
\begin{align}
\Psi^{\pm}(z):= & \im\left[\mathbf{M}_{x_{0},t_{0}}(z)\right]_{2,2}-\im\left[\mathbf{M}_{x_{0},t_{0}}(z)\right]_{1,1}\label{eq:PSI-pm}\\
 & -\sqrt{\pm1}\sqrt{\kappa}(\left[\mathbf{M}_{x_{0},t_{0}}(z)\right]_{2,1}\pm\kappa\left[\mathbf{M}_{x_{0},t_{0}}(z)\right]_{1,2}).\nonumber 
\end{align}

\begin{lem}
\label{lem:aux-spectrum-MaAblo-via-PSI}The auxiliary spectra (\ref{eq:finite-gap-aux-spec1-MaAbl})--(\ref{eq:finite-gap-aux-spec2-MaAbl})
satisfy 
\begin{align}
\left\{ \varrho_{j}(x_{0},t_{0})\right\} _{j=1}^{N}= & \{z\in\mathbb{C}\backslash\mathbb{R}:\Psi^{+}(z)=0\},\label{eq:aux-spec-MaAblo1-via-M}\\
\left\{ \xi_{j}(x_{0},t_{0})\right\} _{j=1}^{N}= & \big\{ z\in\mathbb{C}\backslash\mathbb{R}:\Psi^{-}(z)=0\}.\label{eq:aux-spec-MaAblo2-via-M}
\end{align}
We have $\Psi^{\pm}(z)\in\mathbb{R}$ whenever $\kappa=-1$ and $z\in\mathbb{R}$.\end{lem}
\begin{IEEEproof}
We only discuss $\Psi^{+}$. Lemma \ref{lem:monodromy-symmetry-like-props}
implies $\frac{1}{2}\Psi^{+}=\Imf[\mathbf{M}_{x_{0},t_{0}}]_{1,1}-\im\sqrt{\kappa}\Imf[\mathbf{M}_{x_{0},t_{0}}]_{2,1}$.
Hence, $\Psi^{+}(z)\in\mathbb{R}$ if $\kappa=-1$ and $z\in\mathbb{R}$.
Eq. (\ref{eq:aux-spec-MaAblo1-via-M}) follows with (\ref{eq:monodromy-matrix}). 
\end{IEEEproof}
The next lemma will turn out to be essential at a later point.
\begin{lem}
[\cite{Ma1977}, p. 113f] \label{lem:aux-spec-MaAblo-is-real-if-defocusing}
The auxiliary spectra (\ref{eq:finite-gap-aux-spec1-MaAbl})--(\ref{eq:finite-gap-aux-spec2-MaAbl})
are real in the defocusing case $\kappa=-1$.
\end{lem}

\section{Rational Approximations of the Monodromy Matrix\label{sec:Rational-Approximations-of-M}}

In the previous section, the scattering data has been expressed in
terms of the monodromy matrix. The fast NFT algorithms that will be
given later require a numerically tractable approximation of the monodromy
matrix. Hence, in this section, rational approximations 
\begin{equation}
\mathbf{\hat{M}}(z)=\frac{\mathbf{S}(w)}{d(w)}\approx\mathbf{M}_{x_{0},t_{0}}(z),\qquad w=\varphi^{-1}(z),\label{eq:rational-approx-M}
\end{equation}
of the monodromy matrix are derived given $D$ equidistant samples
of the signal $q$. That is, $\mathbf{S}(w)$ is a matrix-valued polynomial
and $d(w)$ is a scalar-valued polynomial, respectively. The function
$\varphi$ denotes a coordinate transform. Unless specified otherwise,
we shall use a M\"obius transform
\begin{equation}
\varphi^{-1}(z)=\frac{dz-b}{a-cz},\quad\varphi(w)=\frac{\varphi_{1}(w)}{\varphi_{2}(w)}=\frac{aw+b}{cw+d}.\label{eq:Moebius-transf}
\end{equation}
Here, $a,b,c,d\in\mathbb{C}$ with $ad-bc\ne0$. This transform has
no influence on the results with respect to exact arithmetic operations,
but it can be used to improve the numerical properties of the problem
in finite precision. Specific choices will be discussed at the end
of the section.

\subsection{Ansatz}

The monodromy matrix has been defined in terms of the two solutions
to the differential equation in Eq. (\ref{eq:monodromy-matrix}) that
arise from the initial conditions in Eq. (\ref{eq:boundary-conditions-canonical-evs}).
Define the quantity 
\begin{equation}
\boldsymbol{P}_{z}(x):=\left[\begin{array}{cc}
-\im z & -q(x,t_{0})\\
\kappa\bar{q}(x,t_{0}) & \im z
\end{array}\right].\label{eq:P}
\end{equation}
The two solutions can be joined into the single equation 
\begin{equation}
\frac{d}{dx}\mathbf{V}_{z}=\mathbf{P}_{z}\mathbf{V}_{z},\quad\mathbf{V}_{z}(x_{0})=\mathbf{I}.\label{eq:joint-DE-for-M}
\end{equation}
The monodromy matrix can now be written as
\begin{equation}
\mathbf{M}(z)=\mathbf{V}_{z}(x_{0}+\ell).\label{eq:M-via-V}
\end{equation}
The general idea will be to replace the differential equation (\ref{eq:joint-DE-for-M})
with a difference equation which is then solved for an approximation
of $\mathbf{V}_{z}(x_{0}+\ell)$. The difference equation will be
based on given samples of $q(\cdot,t_{0})$ taken at the sample points
\begin{equation}
x_{n}:=x_{0}+n\varepsilon,\quad\varepsilon:=\frac{\ell}{D},\label{eq:sample-points-xn}
\end{equation}
where $n\in\{0,\dots,D-1\}$. Knowing that $\mathbf{P}_{z}(x_{0}+\ell)=\mathbf{P}_{z}(x_{0})$
because $q(\cdot,t_{0})$ is periodic, also set $x_{D}:=x_{0}$.

\subsection{Forward Euler Method}

This method is arguably the simplest way to solve Eq. (\ref{eq:joint-DE-for-M})
for $\mathbf{V}_{z}(x_{D})=\mathbf{M}(z)$. Although it is rarely
used in practice, it is a nice and simple means to illustrate the
general rational approximation in Eq. (\ref{eq:rational-approx-M}).
The discretized version of Eq. (\ref{eq:joint-DE-for-M}) in this
scheme is 
\[
\frac{\hat{\mathbf{V}}_{z}[n+1]-\hat{\mathbf{V}}_{z}[n]}{\varepsilon}=\mathbf{P}_{z}(x_{n})\hat{\mathbf{V}}_{z}[n],\quad\hat{\mathbf{V}}[0]=\mathbf{I}.
\]
Solving for $\hat{\mathbf{V}}_{z}[n+1]$ results in
\begin{equation}
\hat{\mathbf{V}}_{z}[n+1]=\left(\mathbf{I}+\varepsilon\mathbf{P}_{z}(x_{n})\right)\hat{\mathbf{V}}_{z}[n].\label{eq:euler-iteration}
\end{equation}
Eq. (\ref{eq:M-via-V}) suggests the approximation\footnote{Please note the order to the matrix product: $\prod_{k=1}^{K}\mathbf{A}_{k}=\mathbf{A}_{K}\times\dots\times\mathbf{A}_{1}$.}
\begin{align*}
\hat{\mathbf{M}}(z):= & \hat{\mathbf{V}}_{\varphi(w)}[D]\\
= & \prod_{n=1}^{D}\left(\mathbf{I}+\varepsilon\mathbf{P}_{\varphi(w)}(x_{n})\right)\\
= & \prod_{n=1}^{D}\left[\begin{array}{cc}
1-\im\varepsilon\varphi(w) & -\varepsilon q(x_{n},t_{0})\\
\varepsilon\kappa\bar{q}(x_{n},t_{0}) & 1+\im\varepsilon\varphi(w)
\end{array}\right]\\
= & \frac{\prod_{n=1}^{D}\left[\begin{array}{cc}
\varphi_{2}(w)-\im\varepsilon\varphi_{1}(w) & -\varepsilon\varphi_{2}(w)q(x_{n},t_{0})\\
\varepsilon\kappa\varphi_{2}(w)\bar{q}(x_{n},t_{0}) & \varphi_{2}(w)+\im\varepsilon\varphi_{1}(w)
\end{array}\right]}{\varphi_{2}(w)}\\
=: & \frac{\mathbf{S}(w)}{d(w)}.
\end{align*}
At this point, note that $\mathbf{S}(w)$ and $d(w)$ are indeed polynomials.
The coordinate transform $\varphi(w)$ has been incorporated into
them.

\subsection{Crank-Nicolson}

The Crank-Nicolson method is a quite popular finite difference scheme
that is used in practice. The right side of Eq. (\ref{eq:joint-DE-for-M})
is now approximated with a central difference:
\[
\frac{\hat{\mathbf{V}}_{z}[n+1]-\hat{\mathbf{V}}_{z}[n]}{\varepsilon}=\frac{\mathbf{P}_{z}(x_{n+1})\hat{\mathbf{V}}_{z}[n+1]+\mathbf{P}_{z}(x_{n})\hat{\mathbf{V}}_{z}[n]}{2}.
\]
Solving for $\hat{\mathbf{V}}_{z}[n+1]$ results in
\begin{equation}
\hat{\mathbf{V}}_{z}[n+1]=\left(\mathbf{I}-\frac{\varepsilon}{2}\mathbf{P}_{z}(x_{n+1})\right)^{-1}\left(\mathbf{I}+\frac{\varepsilon}{2}\mathbf{P}_{z}(x_{n})\right)\hat{\mathbf{V}}_{z}[n].\label{eq:iteration-CN}
\end{equation}
As before, the monodromy matrix will be approximated using the following
ansatz suggested by Eq. (\ref{eq:M-via-V}): 
\[
\hat{\mathbf{M}}(z):=\hat{\mathbf{V}}_{\varphi(w)}[D]=\hat{\mathbf{V}}_{z}[D],\quad\hat{\mathbf{V}}_{\varphi(w)}[0]:=\mathbf{I}.
\]
The determinant of $\mathbf{I}-\frac{\varepsilon}{2}\mathbf{P}_{\varphi(w)}(x_{n+1})$
is 
\[
\frac{\varphi_{2}^{2}(w)+\frac{\varepsilon^{2}}{4}\left(\varphi_{1}^{2}(w)+\kappa\varphi_{2}^{2}(w)|q(x_{n+1},t_{0})|^{2}\right)}{\varphi_{2}^{2}(w)}=:\frac{d_{n}(w)}{\varphi_{2}^{2}(w)},
\]
and the ansatz expands to
\begin{align*}
\hat{\mathbf{M}}(z)= & \hat{\mathbf{V}}_{\varphi(w)}[D]\\
= & \prod_{n=1}^{D}\left(\mathbf{I}-\frac{\varepsilon}{2}\mathbf{P}_{\varphi(w)}(x_{n+1})\right)^{-1}\left(\mathbf{I}+\frac{\varepsilon}{2}\mathbf{P}_{\varphi(w)}(x_{n})\right)\\
= & \prod_{n=1}^{D}\frac{\varphi_{2}^{2}(w)}{d_{n}(w)}\left[\begin{array}{cc}
1-\im\frac{\varepsilon}{2}\varphi(w) & -\frac{\varepsilon}{2}q(x_{n+1},t_{0})\\
\frac{\varepsilon}{2}\kappa\bar{q}(x_{n+1},t_{0}) & 1+\im\frac{\varepsilon}{2}\varphi(w)
\end{array}\right]\\
 & \times\left[\begin{array}{cc}
1-\im\frac{\varepsilon}{2}\varphi(w) & -\frac{\varepsilon}{2}q(x_{n},t_{0})\\
\frac{\varepsilon}{2}\kappa\bar{q}(x_{n},t_{0}) & 1+\im\frac{\varepsilon}{2}\varphi(w)
\end{array}\right]=\frac{\mathbf{S}(w)}{d(w)},
\end{align*}
where $d(w):=\prod_{n=1}^{D}d_{n}(w)$ and
\begin{align*}
\mathbf{S}(w):= & \prod_{n=1}^{D}\left[\begin{array}{cc}
\varphi_{2}(w)-\im\frac{\varepsilon}{2}\varphi_{1}(w) & -\frac{\varepsilon}{2}\varphi_{2}(w)q(x_{n+1},t_{0})\\
\frac{\varepsilon}{2}\kappa\varphi_{2}(w)\bar{q}(x_{n+1},t_{0}) & \varphi_{2}+\im\frac{\varepsilon}{2}\varphi_{1}(w)
\end{array}\right]\\
 & \times\left[\begin{array}{cc}
\varphi_{2}(w)-\im\frac{\varepsilon}{2}\varphi_{1}(w) & -\frac{\varepsilon}{2}\varphi_{2}(w)q(x_{n},t_{0})\\
\frac{\varepsilon}{2}\kappa\varphi_{2}(w)\bar{q}(x_{n},t_{0}) & \varphi_{2}(w)+\im\frac{\varepsilon}{2}\varphi_{1}(w)
\end{array}\right].
\end{align*}

\subsection{Ablowitz-Ladik Scheme}

The following scheme is known as the Ablowitz-Ladik scheme \cite{Ablowitz1976}:
\begin{align*}
\alpha_{n}:= & \sqrt{1+\kappa\varepsilon^{2}|q(x_{n},t_{0})|^{2}},\\
\hat{\mathbf{V}}_{z}[n+1]= & \alpha_{n}^{-1}\left[\begin{array}{cc}
\eu^{-\im\varepsilon z} & -\varepsilon q(x_{n},t_{0})\\
\varepsilon\kappa\bar{q}(x_{n},t_{0}) & \eu^{\im\varepsilon z}
\end{array}\right]\hat{\mathbf{V}}_{z}[n].
\end{align*}
Note that it is equivalent to the forward Euler method (\ref{eq:euler-iteration})
up to an error of $O(\varepsilon^{2})$ because $\eu^{\mp\im\varepsilon z}=1\mp\im\varepsilon z+O(\varepsilon^{2})$
and $\alpha_{n}=1+O(\varepsilon^{2})$. The coordinate transform
\begin{equation}
w=\varphi^{-1}(z):=\eu^{-\im\varepsilon z},\quad z=\varphi(w)=\frac{\log w}{-\im\varepsilon},\label{eq:ablowitz-ladik-coord-transf}
\end{equation}
leads to the final form of the iteration:
\begin{equation}
\hat{\mathbf{V}}_{w}[n+1]=\alpha_{n}^{-1}\left[\begin{array}{cc}
w & -\varepsilon q(x_{n},t_{0})\\
\varepsilon\kappa\bar{q}(x_{n},t_{0}) & w^{-1}
\end{array}\right]\hat{\mathbf{V}}_{w}[n].\label{eq:iteration-AL}
\end{equation}
The monodromy matrix will be approximated using the ansatz 
\begin{equation}
\hat{\mathbf{M}}(z):=\hat{\mathbf{V}}_{w}[D]\approx\hat{\mathbf{V}}_{z}^{(\text{Euler})}[D],\quad\hat{\mathbf{V}}_{w}[0]=\mathbf{I},\label{eq:ablowitz-ladik-ansatz}
\end{equation}
in which $\hat{\mathbf{V}}_{z}^{(\text{Euler})}[D]$ is given by Eq.
(\ref{eq:euler-iteration}). This fits into the general framework
of Eq. (\ref{eq:rational-approx-M}) if one chooses $d(w):=w^{D}$
and
\[
\mathbf{S}(w):=\prod_{n=1}^{D}\alpha_{n}^{-1}\left[\begin{array}{cc}
w^{2} & -\varepsilon wq(x_{n-1},t_{0})\\
\varepsilon\kappa w\bar{q}(x_{n-1},t_{0}) & 1
\end{array}\right].
\]

\begin{rem}
The normalization by $\alpha_{n}$ is not always given in the literature,
but it has been reported to improve the numerical properties of the
scheme in some cases \cite{Weideman1997}. 
\end{rem}
{}
\begin{rem}
The discretization (\ref{eq:iteration-AL}) is amenable to a discrete
version of the inverse scattering formalism \cite{Ablowitz2004}.
\end{rem}

\subsection{Heuristic for Choosing The Coordinate Transform\label{sub:Heuristics-Coordinate-Transform}}

Many polynomial operations such as root-finding or even simple evaluation
are known to be problematic in finite precision arithmetic. Often
problems arise if the coefficients of a polynomial cover a range that
is too large for the commonly used IEEE 754 double precision floating
point numbers. The problem can become even worse when a polynomial
of a very high degree is evaluated at arguments $x$ with absolute
values $|x|$ that are not close enough to one. In this case, the
powers $|x|^{0},|x|^{1},|x|^{2},|x|^{3},\dots$ will cover a large
range. Consider the following example, which illustrates the difficulties
for the example of polynomial evaluation: $p_{D}(x)=\sum_{d=1}^{D}\frac{10^{-d}}{D}x^{d}$.
It is $p_{D}(10)=1$ for any $D$, but as is illustrated in Tab. \ref{tab:Evaluation-of-PDx}
the numerical evaluation fails spectacularly for larger degrees. 
\begin{table}[t]
\begin{centering}
\begin{tabular}{|c|c|c|c|c|c|}
\hline 
Degree $D$ & $128$ & $256$ & $512$ & $1024$ & $2048$\tabularnewline
\hline 
\hline 
Naive approach & $1$ & $1$ & NaN & NaN & NaN\tabularnewline
\hline 
Horner's method & $1$ & $1$ & $0.60156$ & $0.30078$ & $0.15039$\tabularnewline
\hline 
Reverse Horner & $1$ & $1$ & NaN & NaN & NaN\tabularnewline
\hline 
\end{tabular}
\par\end{centering}

\caption{\label{tab:Evaluation-of-PDx}Numerical evaluation of $P_{D}(x)$
at $x=10$. The exact result is $P_{D}(10)=1$ for any $D$. NaN is
short for ``Not a Number'' (e.g., $0\times\infty=\text{NaN}$ in
IEEE 754). Horner's method is the recommended method for the numerical
evaluation of polynomials if $|x|\ge1$. Otherwise, the reverse Horner's
method should be used \cite{Burrus2012}. That is, evaluate $P_{D}(u)/u$
at $u=x^{-D}$.}
\end{table}

The coordinate transform $z=\varphi(w)$ is a crucial factor in alleviating
such problems when implementing fast NFTs, but finding good transforms
remains a black art for now. Motivated by the issues just discussed,
we wish to find transforms that will map the region of interest close
to the unit circle. The coordinate transform (\ref{eq:ablowitz-ladik-coord-transf})
of the Ablowitz-Ladik scheme achieves this. For the other schemes,
we use the M\"obius transform (\ref{eq:Moebius-transf}) with $a=-M/\varepsilon$,
$b=-a$, $c=\im$, $d=\im$ where $M=1$ for the Euler scheme and
$M=2$ for Crank-Nicolson, respectively. These transforms map the
real line to the unit circle. At the same time, they cancel several
terms in the rational approximations.

\section{Fast Numerical Nonlinear Fourier Transform Based On Finite-Dimensional
Eigenproblems\label{sec:FNSFT-Eigen}}

In this section, a fast numerical NFT will be proposed that is based
on approximating the main and auxiliary spectra through solutions
of finite-dimensional eigenproblems. Its complexity is an order of
magnitude lower than that of similar known algorithms based on matrix
eigenproblems. This section consists of two parts. First, the algorithm
is introduced. Then, it is compared with some other methods from the
literature.

\subsection{Description of the Algorithm\label{sub:FNSFT-Eigen-Alg}}

The input to this algorithm consists of the samples $q(x_{0},t_{0}),\dots,q(x_{D-1},t_{0})$,
where the $x_{n}$ are given in Eq. (\ref{eq:sample-points-xn}).
The user has to decide for a rational approximation $\hat{\mathbf{M}}(z)=\mathbf{S}(w)/d(w)$,
$z=\varphi(w)$, of the monodromy matrix that fits into Eq. (\ref{eq:rational-approx-M}).
Several such schemes have been described in Sec. \ref{sec:Rational-Approximations-of-M}.
The output of the algorithm will be the \emph{numerical main spectrum}
$\tilde{\lambda}_{j}$ and the \emph{numerical auxiliary spectra}
$\tilde{\mu}_{j}(x_{0},t_{0})$, $\tilde{\varrho}_{j}(x_{0},t_{0})$
and $\tilde{\xi}_{j}(x_{0},t_{0})$. The algorithm proceeds as follows.

\subsubsection{Find the Monomial Basis Expansion}

The polynomials $\mathbf{S}(w)$ and $d(w)$ have mostly been given
in product forms $\mathbf{S}(w)=\prod_{n=1}^{D}\mathbf{S}_{n}(w)$
and $d(w)=\prod_{n=1}^{D}d_{n}(w)$, respectively, where the $\mathbf{S}_{n}(w)$
and $d_{n}(w)$ are polynomials of a low degree $K$. However, in
the following, the coefficients of the polynomials $\mathbf{S}(w)$
and $d(w)$ with respect to the usual monomial basis $w^{0},w^{1},w^{2},\dots$
will be required. Algorithm \ref{alg:Fast-product-of-n-polynomials}
is a simple method that can find the coefficients of a polynomial
in product form performing only $O(D\log^{2}D)$ floating point operations
(\emph{flops}). It computes coefficients $\hat{\mathbf{S}}^{(k)}\in\mathbb{C}^{2\times2}$,
$\hat{d}^{(k)}\in\mathbb{C}$ and normalization constants $W_{S},W_{d}\in\mathbb{Z}$
such that 
\begin{equation}
\hat{\mathbf{M}}(w)=\frac{\mathbf{S}(w)}{d(w)}=\frac{2^{W_{S}}\sum_{k=0}^{\deg(\mathbf{S})}\hat{\mathbf{S}}^{(k)}w^{k}}{2^{W_{d}}\sum_{k=0}^{\deg(d)}\hat{d}^{(k)}w^{k}}=:2^{W_{S}-W_{d}}\frac{\hat{\mathbf{S}}(w)}{\hat{d}(w)}.\label{eq:monomial-expansion-S-and-d}
\end{equation}
The normalization constants arise from an effort to avoid overflows
in Algorithm \ref{alg:Fast-product-of-n-polynomials}. They also ensure
that the largest coefficients among the $[\hat{\mathbf{S}}^{(k)}]_{i,j}$
and $\hat{d}^{(k)}$ are of similar magnitude. The basis two was chosen
for the normalization factor because multiplication and division by
two can be carried out without loss of precision in IEEE 754 floating
point numbers. 
\begin{algorithm}[t]
Let us first illustrate the basic idea before the algorithm is given.
The idea is to form the product in a tree-wise fashion, as is illustrated
below for the case $N=8$: 
\[
\begin{array}{ccccccc}
\mathbf{p}_{1}\\
 & {\searrow\atop \nearrow} & \mathbf{p}_{1}\mathbf{p}_{2}\\
\mathbf{p}_{2} &  &  & \searrow\\
 &  &  &  & \mathbf{p}_{1}\mathbf{p}_{2}\mathbf{p}_{3}\mathbf{p}_{4}\\
\mathbf{p}_{3} &  &  & \nearrow\\
 & {\searrow\atop \nearrow} & \mathbf{p}_{3}\mathbf{p}_{4} &  &  & \searrow\\
\mathbf{p}_{4}\\
 &  &  &  &  &  & \mathbf{p}\\
\mathbf{p}_{5}\\
 & {\searrow\atop \nearrow} & \mathbf{p}_{5}\mathbf{p}_{6} &  &  & \nearrow\\
\mathbf{p}_{6} &  &  & \searrow\\
 &  &  &  & \mathbf{p}_{5}\mathbf{p}_{6}\mathbf{p}_{7}\mathbf{p}_{8}\\
\mathbf{p}_{7} &  &  & \nearrow\\
 & {\searrow\atop \nearrow} & \mathbf{p}_{7}\mathbf{p}_{8}\\
\mathbf{p}_{8}
\end{array}
\]
Assume for a moment that $N=2^{n}$. Then, the algorithm will form
$\frac{N}{2^{i}}$ products in the $i$-th level of the tree with
degrees that are less than $2^{i}K$. If the FFT is used to multiply
polynomials fast \cite[p. 204ff]{Kamen2007}, this leads to a complexity
of $O(\frac{N}{2^{i}}2^{i}K\log(2^{i}K))$ flops for level $i$ \cite{Wahls2013b}.
The tree has $O(\log N)$ levels, thus the overall complexity is $O(NK\log^{2}(NK))$.
A detailed implementation that also works for $N\ne2^{n}$ is given
below. Note that the intermediate products are normalized in order
to avoid over-/underflows. 
\begin{lyxlist}{00.00.0000}
\item [{\emph{Input:}}] $N$ polynomials $\mathbf{p}_{1},\dots,\mathbf{p}_{N}$
of degree at most $K$
\item [{\emph{Output:}}] $W$, $\mathbf{p}=2^{W}\mathbf{p}_{1}\mathbf{p}_{2}\times\dots\times\mathbf{p}_{N}$\end{lyxlist}
\begin{itemize}
\item $W\leftarrow0$
\item while $N\ge2$ do:

\begin{itemize}
\item $N_{m}\leftarrow N\mod2$
\item $N\leftarrow\frac{N-N_{m}}{2}+N_{m}$
\item for $n=1,\dots,N-N_{m}$ do:

\begin{itemize}
\item $\mathbf{p}_{n}\leftarrow\mathbf{p}_{2n-1}\mathbf{p}_{2n}$
\item $a\leftarrow\vert\largestcoefficient(\mathbf{p}_{n})\vert$
\item if $a>0$ then $a\leftarrow\left\lfloor \log_{2}a\right\rfloor $
\item $\mathbf{p}_{n}\leftarrow2^{-a}\mathbf{p}_{n}$
\item $W\leftarrow W+a$
\end{itemize}
\item if $N_{m}\ne0$ then: $\mathbf{p}_{N}\leftarrow\mathbf{p}_{2N-1}$
\end{itemize}
\item $\mathbf{p}\leftarrow\mathbf{p}_{1}$
\end{itemize}
\caption{\label{alg:Fast-product-of-n-polynomials}Fast product of $N$ polynomials
$\mathbf{p}_{1},\dots,\mathbf{p}_{N}$ (probably matrix-valued) with
degree at most $K$.}
\end{algorithm}

\subsubsection{Find the Main Spectrum\label{sub:FNSFT-Eigen-Alg-Main-Spectrum}}

Lemma \ref{lem:Main-spectrum-Via-Floquet-Discriminant} suggests to
approximate the main spectrum by the roots of\footnote{We do not check whether the roots are simple. Multiple roots will
be detected by the root-finding algorithm and can be removed in Step
5, if desired.} 
\begin{align}
\hat{\Delta}(z)\pm1:= & \frac{1}{2}\left([\hat{\mathbf{M}}(w)]_{1,1}+[\hat{\mathbf{M}}(w)]_{2,2}\right)\pm1\nonumber \\
= & 2^{W_{S}-1}\frac{[\hat{\mathbf{S}}(w)]_{1,1}+[\hat{\mathbf{S}}(w)]_{2,2}\pm2^{W_{d}-W_{S}+1}\hat{d}(w)}{\hat{d}(w)}.\label{eq:Numerical-approx-Floquet-pm1}
\end{align}
The roots of these rational functions correspond to the roots of the
two numerators $[\hat{\mathbf{S}}(w)]_{1,1}+[\hat{\mathbf{S}}(w)]_{2,2}\pm2^{W_{d}-W_{S}+1}\hat{d}(w)$
that are not canceled by a root of the denominator $\hat{d}(w)$.
It is a well-known fact that the roots of a polynomial correspond
to the eigenvalues of an associated companion matrix that can be constructed
from its monomial basis expansion. Companion matrices are highly structured,
and recently several algorithms have been proposed that can find the
eigenvalues of an $R\times R$ companion matrix with only $O(R^{2})$
flops. See, e.g., \cite{Chandrasekaran2007,VanBarel2010,Boito2012a,Bevilacqua2014}
and the references therein. We propose to find the roots of the two
polynomials $[\hat{\mathbf{S}}(w)]_{1,1}+[\hat{\mathbf{S}}(w)]_{2,2}\pm2^{W_{d}-W_{S}+1}\hat{d}(w)$
using this method. Therefore, one requires the monomial basis expansion
of these two polynomials. Since the expansion (\ref{eq:monomial-expansion-S-and-d})
is already known, it can be computed in only $O(KD)$ flops. The roots
of $\hat{d}(w)$ can often be found in closed form, or otherwise,
if a product expansion $\hat{d}(w)=\prod_{n=1}^{D}\hat{d}_{n}(w)$
is known, numerically using at most $O(K^{2}D)$ flops. Denote the
roots of the two numerators that are not being canceled by a root
of $\hat{d}(w)$ by $\hat{w}_{j}$. The worst-case complexity of finding
them is $O(K^{2}D^{2})$ because for each root of the numerators one
has to check whether that root is in the set of roots of the denominator.
Often, this step can be simplified when $\hat{d}(w)$ has only a few
distinct roots. We finally apply the coordinate transform in order
to find the numerical main spectrum, $\hat{\lambda}_{j}:=\varphi(\hat{w}_{j})$.
Adding up the complexities of the single steps, we see that the overall
complexity of finding the numerical main spectrum is $O(K^{2}D^{2})$
flops.

\subsubsection{Find the Kotlyarov-Its Auxiliary Spectrum}

The replacement of the monodromy matrix in Lemma \ref{lem:aux-spectrum-via-monodromy}
with the rational approximation (\ref{eq:monomial-expansion-S-and-d})
results in\footnote{We do not check whether the found roots belong to the degenerated
modes at this point. Such roots can be removed later in Step 5, if
desired.} 
\begin{equation}
[\hat{\mathbf{M}}(w)]_{1,2}=2^{W_{S}-W_{d}}\frac{[\hat{\mathbf{S}}(w)]_{1,2}}{\hat{d}(w)}=0.\label{eq:Numerical-approx-M12}
\end{equation}
The roots in this equation are an approximation of the auxiliary spectrum.
They correspond to the roots of the numerator $[\hat{\mathbf{S}}(w)]_{1,2}$
that are not canceled by the roots of $\hat{d}(w)$. Denote the remaining
roots of $[\hat{\mathbf{S}}(w)]_{1,2}$ by $\hat{w}_{j}$. The numerical
auxiliary spectrum (of Kotlyarov-Its type) is given by $\hat{\mu}_{j}(x_{0},t_{0}):=\varphi(\hat{w}_{j})$.
If the same root finder as in Sec. \ref{sub:FNSFT-Eigen-Alg-Main-Spectrum}
is used, it can be computed using $O(K^{2}D^{2})$ flops.

\subsubsection{Find the Ma-Ablowitz Auxiliary Spectra}

Now, the idea is to exploit Lemma \ref{lem:aux-spectrum-MaAblo-via-PSI}.
The replacement of the monodromy matrix in Eq. (\ref{eq:PSI-pm})
with (\ref{eq:monomial-expansion-S-and-d}) results in
\begin{align}
 & \hat{\Psi}^{\pm}(z):=2^{W_{S}-W_{d}}\times\label{eq:PSI-pm-approx}\\
 & \frac{\im[\hat{\mathbf{S}}(w)]_{2,2}-\im[\hat{\mathbf{S}}(z)]_{1,1}-\sqrt{\pm1}\sqrt{\kappa}([\hat{\mathbf{S}}(w)]_{2,1}\pm\kappa[\hat{\mathbf{S}}(w)]_{1,2})}{\hat{d}(w)}.\nonumber 
\end{align}
The roots $\hat{w}_{j}^{\pm}$ of the numerators of $\hat{\Psi}^{\pm}$
that are not canceled by the denominators define the numerical auxiliary
spectra (of Ma-Ablowitz type) $\hat{\varrho}_{j}(x_{0},t_{0}):=\varphi(\hat{w}_{j}^{+})$
and $\hat{\xi}_{j}(x_{0},t_{0}):=\varphi(\hat{w}_{j}^{-})$. The auxiliary
spectra can be computed in $O(K^{2}D^{2})$ flops if the same root
finder as in Sec. \ref{sub:FNSFT-Eigen-Alg-Main-Spectrum} is used.

\subsubsection{Filter the Numerical Spectra}

The numerical spectra will often contain artifacts that arise because
of the discretization procedure. These artifacts will usually be well-separated
from the real spectrum. Whenever there is some a-priori knowledge
about the spectrum, it should be used to remove all other points in
the numerical spectra that contradict this knowledge. The detected
main spectrum may contain double roots indicating degenerate points
in the main spectrum. These can be removed if desired. If a pair of
degenerate points is removed from the main spectrum, then the same
point should also be removed from the auxiliary spectra in order to
remove the corresponding degenerate hyperelliptic modes as well.
\begin{rem}
[Root Cancellations] \label{rem:root-cancellation-may-not-be-necessary}
Depending on the discretization scheme, it is not always necessary
to perform the root cancellations in the algorithm. The discretization
schemes discussed in Sec. \ref{sec:Rational-Approximations-of-M}
lead to denominators $\hat{d}(w)$ with roots that are usually well
separated from the spectrum. The denominator in the Ablowitz-Ladik
scheme has no finite roots at all. In the Euler and Crank-Nicolson
schemes, when used with the M\"obius transforms from Sec. \ref{sub:Heuristics-Coordinate-Transform},
the roots of $\hat{d}(w)$ cluster around $w=-1$ as the step-size
$\varepsilon$ becomes small. In original coordinates, they will cluster
around $z=\varphi(1)=\infty$. The root cancellation steps will therefore
usually not be necessary with the discretization schemes from Sec.
\ref{sec:Rational-Approximations-of-M}.
\end{rem}

\subsection{Comparison With Other Finite-Dimensional Eigenmethods\label{sub:Comparison-With-Other-Eigenmethods}}

Another way to compute the scattering data is by direct discretization
of the eigenrelation $\mathbf{L}_{t_{0}}\mathbf{v}=z\mathbf{v}$.
Let us illustrate this approach with an example. The main spectrum
consists of the eigenvalues of $\mathbf{L}_{t_{0}}$ that possess
(anti-)periodic eigenvectors. The eigenrelation 
\begin{equation}
\mathbf{L}_{t_{0}}\left[\begin{array}{c}
u\\
v
\end{array}\right]=\im\left[\begin{array}{cc}
\frac{d}{dx} & q(\cdot,t_{0})\\
\kappa\bar{q}(\cdot,t_{0}) & -\frac{d}{dx}
\end{array}\right]\left[\begin{array}{c}
u\\
v
\end{array}\right]=z\left[\begin{array}{c}
u\\
v
\end{array}\right]\label{eq:Lv-is-vz}
\end{equation}
can be discretized using Euler's method. With the sample points $x_{n}$
and step size $\varepsilon$ given in Eq. (\ref{eq:sample-points-xn}),
this becomes 
\begin{align}
\im\frac{u(x_{n+1})-u(x_{n})}{\varepsilon}+\im q(x_{n},t_{0})v(x_{n}) & =zu(x_{n}),\label{eq:discretization-of-L-eigenprob-1}\\
\im\kappa\bar{q}(x_{n},t_{0})u(x_{n})-\im\frac{v(x_{n+1})-v(x_{n})}{\varepsilon} & =zv(x_{n}),\label{eq:discretization-of-L-eigenprob-2}
\end{align}
where $n\in\{0,\dots,D-1\}$. The signal $q$ is periodic in $x$
and the eigenfunctions are supposed to be (anti-)periodic. Hence,
$q(x_{D},t_{0})=q(x_{0},t_{0})$ and $u(x_{D})=\pm u(x_{0})$, $v(x_{D})=\pm v(x_{0})$.
The $2D$ linear equations in Eqs. (\ref{eq:discretization-of-L-eigenprob-1})--(\ref{eq:discretization-of-L-eigenprob-2})
can be collected into a single (generalized) matrix eigenproblem that
can be solved with standard numerical methods such as the QZ algorithm.
Each sign $\pm$ results in a separate eigenproblem, and the collected
eigenvalues of both problems can be used as an approximation of the
true main spectrum. The same approach can be used to approximate auxiliary
spectra if the boundary conditions are modified appropriately. See,
e.g., \cite[Ch. 3.4]{MacEvoy1994}. Of course, methods other than
Euler's can be used in order to discretize Eq. (\ref{eq:Lv-is-vz})
including such where Eq. (\ref{eq:Lv-is-vz}) is discretized in the
Fourier domain. Several such methods have been proposed in the literature
\cite{Yousefi2012b,Weideman1997,Burtsev1998,ChamorroPosada1999},
but so far no way of exploiting the structure of the resulting matrices
seems to be known. Instead, standard eigenvalue solvers with $O(D^{3})$
complexity had to be used. This is in contrast to our proposed algorithm,
which requires only $O(D^{2})$ flops.
\begin{rem}
[Floquet-Fourier-Hill Method]Deconinck and Kutz \cite{Deconinck2006}
proposed a novel approach to the numerical approximation of the spectrum
of periodic differential operators like $\mathbf{L}_{t_{0}}$. Although
their method is quite different from the ones discussed above, it
still has a complexity of $O(D^{3})$.
\end{rem}

\subsection{Comparison With Search Methods\label{sub:Comparison-With-Search-Methods}}

Search methods are among the oldest methods for computing the nonlinear
Fourier transform \cite{Boffetta1992}. Let us illustrate the basic
idea for the main spectrum. As before, the roots of the function $\hat{\Delta}(z)\pm1$
will be used as the numerical approximations of the main spectrum.
The difference is that this time, an iterative search method such
as Newton's method will be used in order to locate the roots. Let
$r[0]$ denote an estimate for a root of $\hat{\Delta}(z)\pm1$. The
derivative with respect to a complex variable $z=u+\im v$, $u,v\in\mathbb{R}$,
is $\frac{d}{dz}:=\frac{1}{2}\left(\partial_{u}-\im\partial_{v}\right)$.
Therewith, the complex Newton's method can be written as \cite{Yau1998}
\begin{align}
r[i+1]= & r[i]+\left.\frac{\hat{\Delta}\pm1}{\frac{d}{dz}(\hat{\Delta}\pm1)}\right|_{z=r[n]}\nonumber \\
= & r[i]+\left.\frac{[\hat{\mathbf{M}}]_{1,1}+[\hat{\mathbf{M}}]_{2,2}\pm2}{[\frac{d}{dz}\hat{\mathbf{M}}]_{1,1}+[\frac{d}{dz}\hat{\mathbf{M}}]_{2,2}}\right|_{z=r[i]}.\label{eq:Newton-step}
\end{align}
The derivative of the approximated monodromy matrix $\hat{\mathbf{M}}(z)$
is required in order to perform this iteration. The exact formulas
for this derivative depend on the discretization method. For Euler's
method, e.g., it is $\hat{\mathbf{M}}(z)=\hat{\mathbf{V}}_{z}[D]$
where $\hat{\mathbf{V}}[D]$ is determined through the iteration (\ref{eq:euler-iteration}).
Consequently, $\frac{d}{dz}\hat{\mathbf{M}}(z)=\frac{d}{dz}\hat{\mathbf{V}}_{z}[D]$.
Taking the derivative of (\ref{eq:euler-iteration}) results in 
\begin{align}
\frac{d}{dz}\hat{\mathbf{V}}_{z}[n+1]= & \left(\mathbf{I}+\varepsilon\mathbf{P}_{z}(x_{n})\right)\frac{d}{dz}\hat{\mathbf{V}}_{z}[n]\nonumber \\
 & +\varepsilon\left(\frac{d}{dz}\mathbf{P}_{z}(x_{n})\right)\hat{\mathbf{V}}_{z}[n].\label{eq:euler-iteration-deriv}
\end{align}
Differentiation of the initial condition $\hat{\mathbf{V}}_{z}[0]=\mathbf{I}$
gives the missing initial condition for the derivative: $\frac{d}{dz}\hat{\mathbf{V}}_{z}[0]=\mathbf{0}.$

In summary, given an initial guess $r[0]$ of a root of $\hat{\Delta}\pm1$,
one iterates the Newton step (\ref{eq:Newton-step}) until some stopping
criterion is fulfilled. For example, one may stop whenever the difference
between two consecutive iterates falls below a certain threshold \cite{Yousefi2012b}.
That is, $\vert r[i+1]-r[i]\vert\le\beta$ for some $\beta>0$. The
monodromy matrix and its derivative, which are required for each Newton
step, can be found by using (\ref{eq:euler-iteration}) and (\ref{eq:euler-iteration-deriv}).

There are two remaining issues that have to be solved for any search
method:
\begin{enumerate}
\item How does one choose the initial guesses for the roots?
\item How does one know that all roots of interest have been found? Newton's
method does not return the multiplicities of the found roots.
\end{enumerate}
These issues have been discussed in \cite{Yousefi2012b} for the NFT
with non-periodic signals that vanish at infinity. In the communication
scenario of \cite{Yousefi2013}, in which there are only finitely
many possible locations for the roots, solving these problems is simple.
It was proposed in \cite{Yousefi2012b} to use a few random perturbations
of each potential root as initial guesses. Since Newton's method converges
quickly for initial guesses that are close to a potential root, one
can assume that all roots have been found after each potential root
has been tested. The cost of performing one Newton step (\ref{eq:Newton-step})
using the iterations (\ref{eq:euler-iteration}) and (\ref{eq:euler-iteration-deriv})
is $O(D)$ flops. The number of iterations per root can be as high
as $O(D)$ as well \cite[p. 936]{Schleicher2002}. Thus, in a communication
scenario with $P$ possible values for the spectral points, the worst-case
complexity is at least $O(PD^{2})$ flops. (The average complexity
usually will be better, though.)

The situation becomes more involved in situations without precise
a-priori knowledge. It was proposed in \cite{Yousefi2012b} to define
a fixed region in which the spectrum is expected to be, and to choose
initial guesses for the roots uniformly at random from this region.
Regarding the stopping criterion, the authors of \cite{Yousefi2012b}
proposed to check whether the spectrum found so far satisfies an energy
conservation law. In the non-periodic case, the energy conservation
law only involves the evaluation of an integral over the real line
(see, e.g., \cite[Apdx. B]{Yousefi2012a}). The corresponding law
for the periodic case however involves other roots that also have
to be found \cite[Eq. (4.27)]{Tracy1991}. Hence, this stopping criterion
seems to be less appropriate in the periodic case. Even very recent
results on the initialization of Newton's method like \cite{Hubbard2001}
and \cite{Bilarev2012} do not achieve $O(D^{2})$ complexity.

In summary, search methods seem to be a good choice only if there
is precise a-priori knowledge, and the number of roots $P$ is low
compared to the number of sample points $D$.

\section{Fast Numerical Nonlinear Fourier Transform For The Defocusing NSE
Based On Sampling\label{sec:FNFT-Sampling}}

In this section, an especially fast numerical NFT that finds the main
spectrum as well as the Ma-Ablowitz auxiliary spectra for the defocusing
NSE (\ref{eq:NSE}) will be proposed. The idea is to exploit the fact
that these spectra are always real in the defocusing case. The Kotlyarov-Its
auxiliary spectrum does not have to be real and cannot be found with
the method described in the following. The section is structured as
follows. First, the algorithm is described. Then, its connection to
a related approach from the literature is discussed.

\subsection{Description of the Algorithm\label{sub:FNFT-Samp-Alg}}

The inputs and outputs are the same as in Sec. \ref{sub:FNSFT-Eigen-Alg}
with the exception that additionally a lower bound $A\in\mathbb{R}$
and an upper bound $B\in\mathbb{R}$, $A<B$, on the spectra have
to be provided.

\subsubsection{Find the Monomial Basis Expansion}

One starts as in Sec. \ref{sub:FNSFT-Eigen-Alg}. First, the monomial
basis expansions in Eq. (\ref{eq:monomial-expansion-S-and-d}) are
computed. All other expansions that will be required in this algorithm
can be found from this information with $O(D)$ flops.

\subsubsection{Find the Main Spectrum\label{sub:FNFT-Samp-Alg-Main-Spectrum}}

As in Sec. \ref{sub:FNSFT-Eigen-Alg}, the numerical main spectrum
will be found from the roots of (\ref{eq:Numerical-approx-Floquet-pm1}).
However, this time Lemma \ref{lem:monodromy-symmetry-like-props}
ensures that $\Delta(z)\pm1$ is real-valued on the real axis. Additionally,
all roots are real by Lemma \ref{lem:main-spectrum-symmetry}. This
makes the root finding problem much easier. Algorithm \ref{alg:NFT-on-the-line}
shows a straight-forward three-step method to isolate the roots. First,
$\hat{\Delta}(z)$ is sampled on an equidistant grid. Second, the
adjacent sample points where any of the two functions $\hat{\Delta}(z)\pm1$
changes its sign are located. The midpoint of any such pair of adjacent
sample points forms an estimate of a root. In the third step, these
estimates are refined using bisection.

\begin{algorithm}[t]
\begin{lyxlist}{00.00.0000}
\item [{Input:}] Bounds $A,B\in\mathbb{R}$, oversampling factor $G\in\mathbb{N}$,
number of bisection iterations $L\in\mathbb{N}$
\item [{Output:}] Numerical main spectrum $\tilde{\lambda}_{1},\dots,\tilde{\lambda}_{R}$\end{lyxlist}
\begin{itemize}
\item for $n=0,\dots,GD-1$: $w_{n}=\varphi^{-1}(A+n\frac{B-A}{GD-1})$,
\[
v_{n}\leftarrow\hat{\Delta}(\varphi(w_{n}))=2^{W_{S}-W_{d}-1}\frac{[\hat{\mathbf{S}}(w_{n})]_{1,1}+[\hat{\mathbf{S}}(w_{n})]_{2,2}}{\hat{d}(w_{n})}
\]

\item $R\leftarrow-1$
\item for $n=0,\dots,GD-2$:

\begin{itemize}
\item if $\sign(\Re(v_{n+1})-1))\ne\sign(\Re(v_{n})-1)$:

\begin{itemize}
\item $R\leftarrow R+1$, $a_{R}\leftarrow z_{n}$, $b_{R}\leftarrow z_{n+1}$
\item $s_{R}\leftarrow+1$, $u_{R}\leftarrow v_{n}$
\end{itemize}
\item if $\sign(\Re(v_{n+1})+1)\ne\sign(\Re(v_{n})+1)$:

\begin{itemize}
\item $R\leftarrow R+1$, $a_{R}\leftarrow z_{n}$, $b_{R}\leftarrow z_{n+1}$
\item $s_{R}\leftarrow-1$, $u_{R}\leftarrow v_{n}$
\end{itemize}
\end{itemize}
\item for $l=1,\dots,L$:

\begin{itemize}
\item for $i=1,\dots,R$: $u_{r}\leftarrow\hat{\Delta}(\varphi^{-1}(\frac{a_{r}+b_{r}}{2}))$
\item for $i=1,\dots,R$:

\begin{itemize}
\item if $\sign(\Re(u_{r})-s_{r})=\sign(\Re(v_{r})-s_{r})$:\\
\phantom{asd}$a_{r}\leftarrow\frac{a_{r}+b_{r}}{2}$, $u_{r}\leftarrow v_{r}$
\item else: $b_{r}\leftarrow\frac{a_{r}+b_{r}}{2}$
\end{itemize}
\end{itemize}
\item for $i=1,\dots,R$: $\tilde{\lambda}_{r}\leftarrow\frac{a_{r}+b_{r}}{2}$
\end{itemize}
\caption{\label{alg:NFT-on-the-line}The root finding procedure for Sec.\ref{sub:FNFT-Samp-Alg-Main-Spectrum}.
The evaluations of $\hat{\Delta}(z)$ have to implemented using the
non-equidistant Fourier transform \cite{Keiner2009} in order to obtain
an $O(LGD\log(GD))$ algorithm.}
\end{algorithm}

The costs of Algorithm \ref{alg:NFT-on-the-line} are dominated by
the evaluations of $\hat{\Delta}(z)$. With naive evaluations of (\ref{eq:iteration-CN}),
(\ref{eq:iteration-AL}), etc., the overall costs of the algorithm
are $O(LG^{2}D^{2})$. However, note that the coordinate transforms
$w=\varphi^{-1}(z)$ in Sec. \ref{sub:Heuristics-Coordinate-Transform}
as well as the transform (\ref{eq:ablowitz-ladik-coord-transf}) map
the real axis to the complex unit circle. Therefore, the \emph{non-equidistant
fast Fourier transform (NFFT)} \cite{Keiner2009} can be used in order
to evaluate $\hat{\Delta}(z)$ efficiently. For example, the $v_{0},\dots,v_{GD-1}$
can be found as follows. First, one uses the NFFT to evaluate $[\hat{\mathbf{S}}(w)]_{1,1}+[\hat{\mathbf{S}}(w)]_{2,2}$
at the $GD$ points $w_{n}:=\varphi^{-1}(z_{n})$ using only $O(GD\log(GD))$
flops because $|w_{n}|=1$ for all $n$. Second, one uses the NFFT
to evaluate $\hat{d}(w)$ at the $w_{n}$. Finally, the values of
$\hat{\Delta}(z)$ are given by 
\[
v_{n}=2^{W_{S}-W_{d}-1}\frac{[\hat{\mathbf{S}}(w_{n})]_{1,1}+[\hat{\mathbf{S}}(w_{n})]_{2,2}}{\hat{d}(w_{n})}.
\]
Both NFFTs require $O(GD\log(GD))$ flops. The last step requires
another $O(GD)$ flops. In total, one can thus find $v_{0},\dots,v_{GD-1}$
using only $O(GD\log(GD))$ flops. Similarly, the $u_{1},\dots u_{R}$
can be found using only $O(GD\log(GD))$ flops because $R\le GD$.
The costs of the remaining operations are negligible, so that the
overall costs of Algorithm \ref{alg:NFT-on-the-line} amount to $O(LGD\log(GD))$
flops if the NFFT is utilized.

\subsubsection{Find the Ma-Ablowitz Auxiliary Spectra}

The roots of the function $\Psi^{\pm}(z)$ defined in (\ref{eq:PSI-pm})
constitute the auxiliary spectra. These roots are real by Lemma \ref{lem:aux-spec-MaAblo-is-real-if-defocusing}.
Furthermore, Lemma \ref{lem:aux-spectrum-MaAblo-via-PSI} shows that
$\Psi^{\pm}(z)$ is real for real $z$. The auxiliary spectra can
therefore be found with the same method as the main spectrum, using
only $O(LGD\log(GD))$ flops. One only has to replace $\hat{\Delta}\pm1$
with $\hat{\Psi}^{\pm}$ as defined in Eq. (\ref{eq:PSI-pm-approx}).

\subsubsection{Filter the Numerical Spectra}

As in Sec. \ref{sub:FNSFT-Eigen-Alg}. The spectra of the defocusing
NSE have a special band structure \cite[p. 117]{Ma1981}, which can
help to identify numerical artifacts.

\begin{rem}
[Root Refinements] In Algorithm \ref{alg:NFT-on-the-line}, bisection
has been used to refine the roots. Of course, more advanced methods
like the secant method, Muller's method \cite{Muller1956}, or Ridders'
method \cite{Ridders1979} could be used as well. Since derivatives
are easily found in the Fourier domain, the NFFT can also be used
the compute the derivative of $\hat{\Delta}(z)$ efficiently. Thus,
even the use of Newton's method should be possible. 
\end{rem}
{}
\begin{rem}
[Root Cancellations] In contrast to the fast eigenmethod from Sec.
\ref{sub:FNSFT-Eigen-Alg}, the algorithm in this section did not
contain any root cancellation procedures. Root-finding methods based
on sampling do not indicate the multiplicities of the found roots.
Hence, one cannot detect, e.g., whether a root of $[\hat{\mathbf{S}}(w)]_{1,1}+[\hat{\mathbf{S}}(w)]_{2,2}$
will been canceled by a root of $\hat{d}(w)$ unless it is assumed
that the roots of $[\hat{\mathbf{S}}(w)]_{1,1}+[\hat{\mathbf{S}}(w)]_{2,2}$
are simple. Fortunately, Remark \ref{rem:root-cancellation-may-not-be-necessary}
applies in this case as well.
\end{rem}

\subsection{Comparison With A Similar Approach}

Algorithm \ref{alg:NFT-on-the-line} is quite similar to a numerical
NFT proposed by Christov in the context of the Korteweg-de Vries equation
\cite{Christov2009}, whose associated Lax operator has a real spectrum
as well. There are some differences in terms of the root refinement,
but the main difference to our approach, which actually is the key
to obtaining a fast algorithm, however is that rational approximations
of the monodromy matrix are used in this paper. In contrast, an irrational
approximation of the monodromy matrix has been used in \cite{Christov2009}.
Since an evaluation of the monodromy matrix in \cite{Christov2009}
takes $O(GD)$ flops, the overall runtime of the algorithm in \cite{Christov2009}
is about $O(LD^{2}G^{2})$ flops. This is about an order of magnitude
higher than for our implementation, which has runtime of $O(LGD\log(GD)+KD\log^{2}(KD))$.
Here, the second term is due to Algorithm \ref{alg:Fast-product-of-n-polynomials}.

\section{Numerical Examples\label{sec:Numerical-Examples}}

In this section, two numerical examples are investigated in order
to demonstrate that the new proposed algorithms can improve significantly
over the existing ones in terms of runtimes without sacrificing numerical
reliability. The first example evaluates the finite eigenmethod from
Sec. \ref{sub:FNSFT-Eigen-Alg}, while the second example investigates
the performance of the specialized method from Sec. \ref{sub:FNFT-Samp-Alg}
for the defocusing NSE. The results presented here should be understood
as proof-of-concept. A comprehensive numerical or analytical study
of the new algorithms remains a topic for future research.\emph{ }
\begin{rem}
The numerical examples have been carried out using MATLAB R2012a,
but for the most time-consuming parts of the algorithms external implementations
have been used. In order to realize the root-finding algorithm in
Sec. \ref{sub:FNSFT-Eigen-Alg-Main-Spectrum}, a Fortran implementation
of the algorithm in \cite{Boito2012a}, which is available at \url{http://www.unilim.fr/pages_perso/paola.boito/software.html},
has been interfaced. For the non-equidistant fast Fourier transform
required in Sec. \ref{sub:FNFT-Samp-Alg-Main-Spectrum}, version 3.2.3
of the NFFT3 library \cite{Keiner2009}, which is available at \url{http://www.tu-chemnitz.de/~potts/nfft},
has been used. Specifically, the included example given in the file
\texttt{test\_nfft1d.m} has been adapted to our needs. Algorithm \ref{alg:Fast-product-of-n-polynomials}
and the direct evaluation of (\ref{eq:iteration-AL}) have been implemented
in C. The FFT in Algorithm \ref{alg:Fast-product-of-n-polynomials}
has been realized using version 1.3.0 of the KISS FFT routine, which
is available at \url{http://sourceforge.net/projects/kissfft/}. Root
cancellations have not been implemented.
\end{rem}

\subsection{The Focusing Case\label{sub:Numerical-Example-Focusing}}

The first example is the initial condition $q(x,t_{0})=q_{0}\eu^{\im\mu x}$,
which has been discussed e.g. in \cite{Weideman1997}. The main spectrum
with respect to the focusing NSE is $\zeta_{n}^{\pm}=-\frac{\mu}{2}\pm\im\sqrt{|q_{0}|^{2}-\frac{n^{2}}{4}}$,
$n\in\mathbb{N}$, where all eigenvalues except $\zeta_{0}^{\pm}$
are double. Similar to Example \ref{exa:plane-wave}, there is thus
one non-degenerate band. 
\begin{figure*}[t]
\begin{centering}
\includegraphics[width=0.45\linewidth]{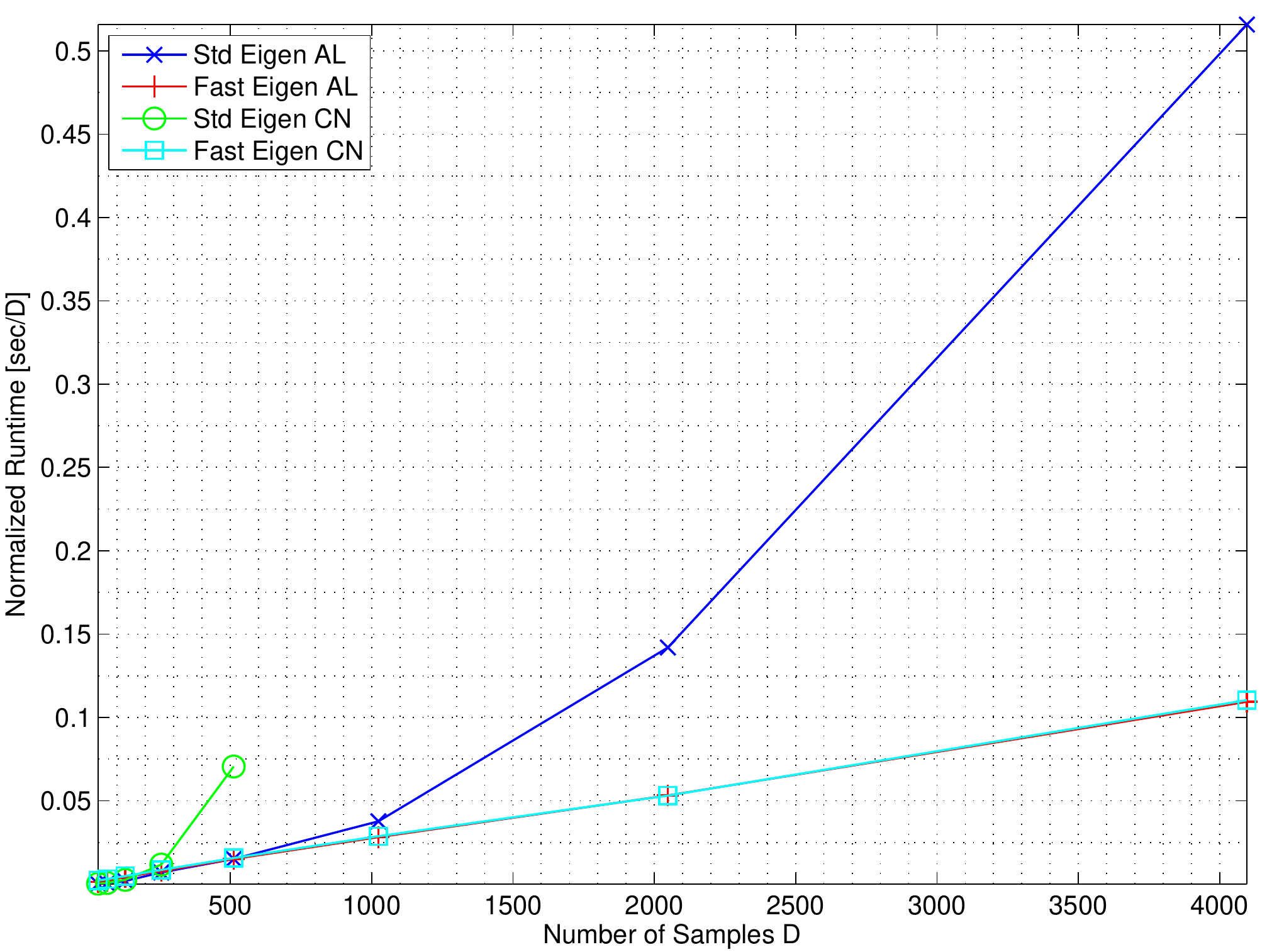}\qquad{}\includegraphics[width=0.45\linewidth]{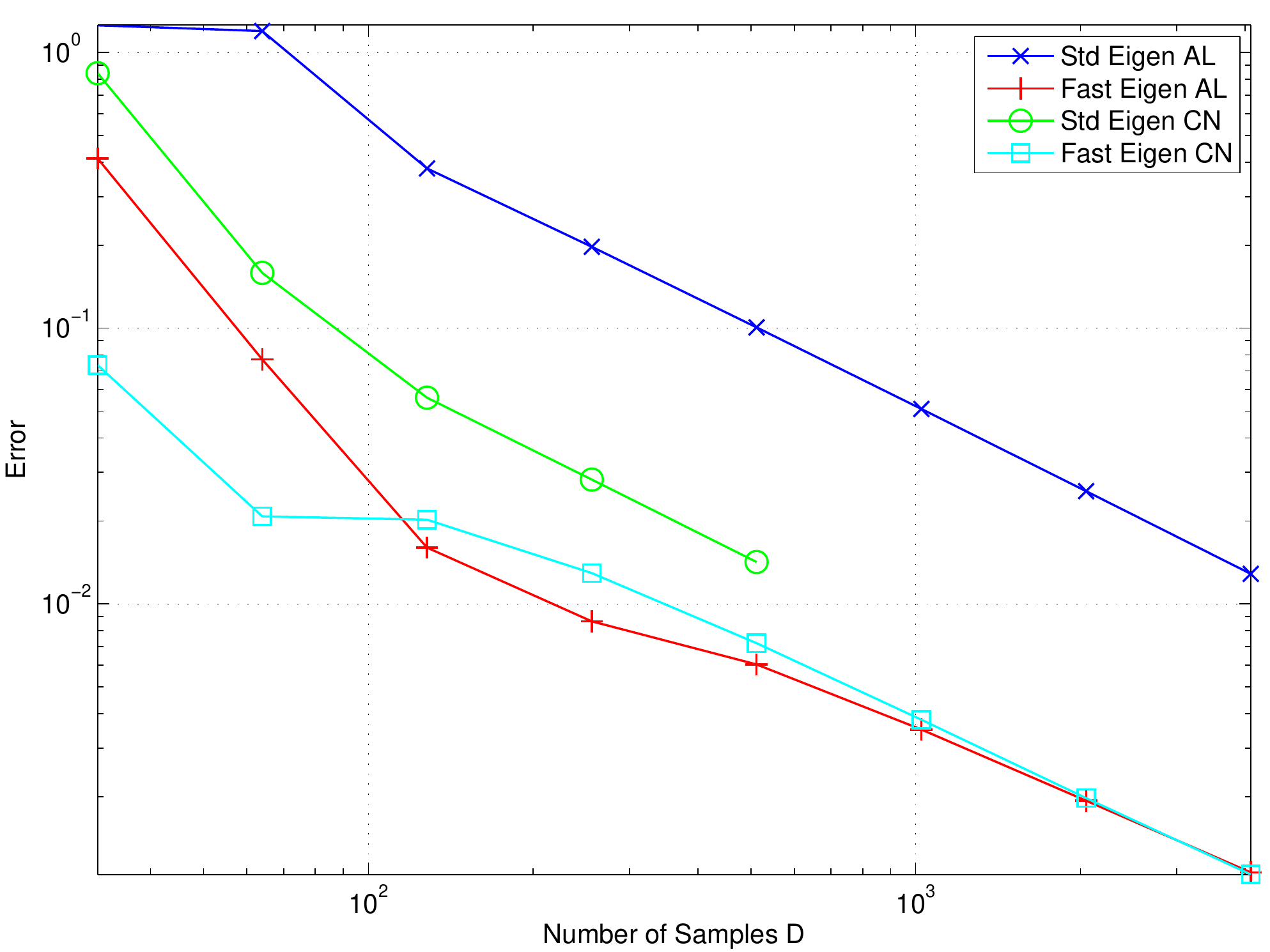}
\par\end{centering}

\centering{}\caption{\label{fig:Runtimes_focusing}Focusing one-band solution. Left: Runtimes
per sample, Right: Error. Please note that the Std-CN method was too
slow for higher $D$.}
\end{figure*}
\begin{figure*}[t]
\centering{}\includegraphics[width=0.45\linewidth]{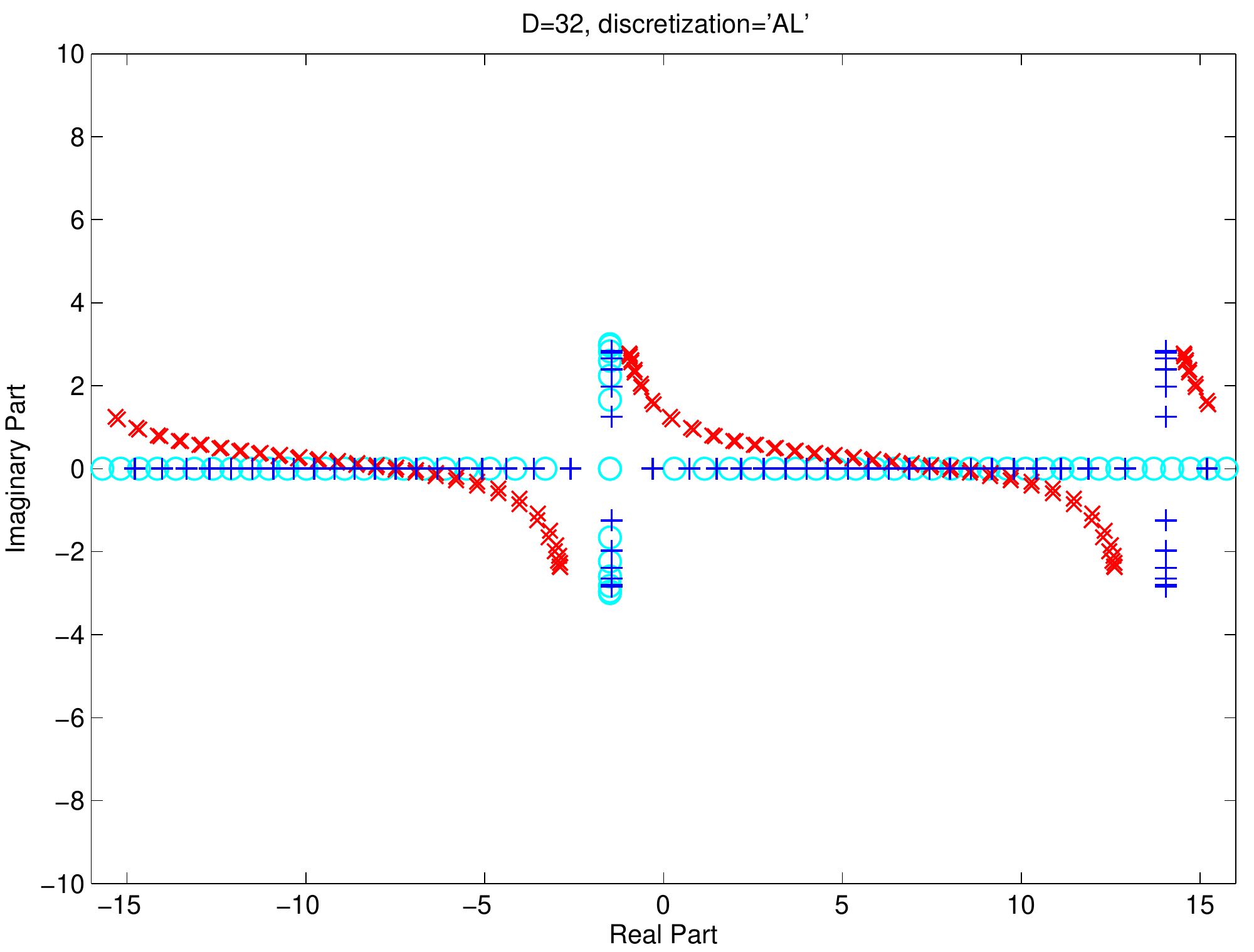}\qquad{}\includegraphics[width=0.45\linewidth]{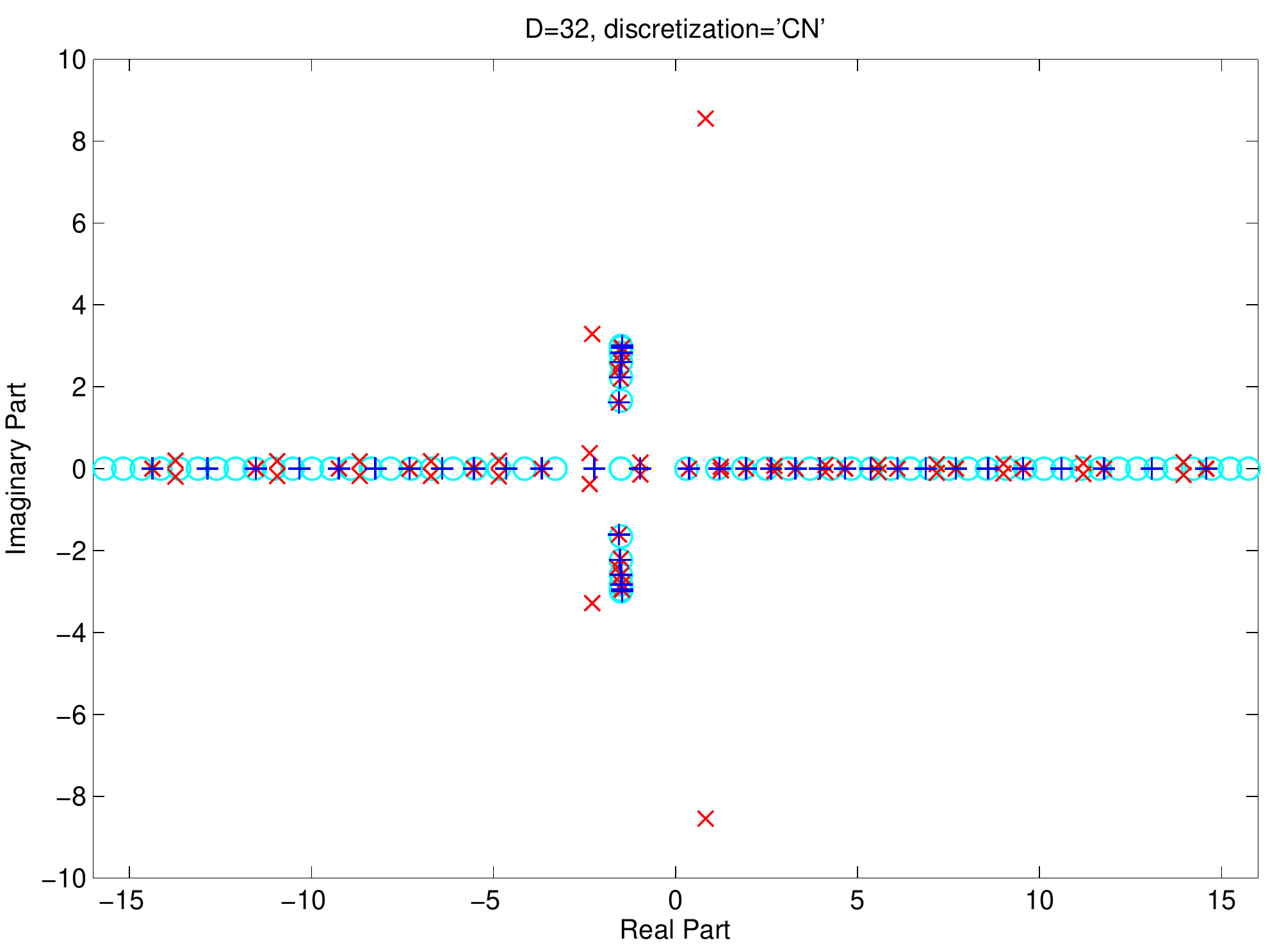}\caption{\label{fig:spectra}Exact main spectrum (\textcolor{cyan}{o}) vs.
numerical main spectra found by the the new fast eigenmethod (\textcolor{blue}{+})
and the conventional eigenmethod (\textcolor{red}{x}) }
\end{figure*}

First, the runtimes of the new fast numerical NFT from Sec. \ref{sub:FNSFT-Eigen-Alg}
will be compared with finite difference methods as described in Sec.
\ref{sub:Comparison-With-Other-Eigenmethods}. Two discretizations
will be considered: the Ablowitz-Ladik (``AL'') discretization and
the Crank-Nicolson (``CN'') discretization. The signal parameters
are $q_{0}=\mu=3$ and $\ell=2\pi$. The first plot in Fig. \ref{fig:Runtimes_focusing}
shows the minimum runtime \emph{per sample point}, taken over three
runs, versus the number of sample points $D$. The per-sample runtimes
of the fast algorithms grow approximately linearly with $D$ as expected,
while the per-sample runtimes of the standard algorithms grow quadratically.
This corresponds to quadratic and cubic overall runtimes, respectively. 

Next, the numerical accuracy of these algorithm is compared. Since
the signal has infinitely many degenerate modes, only the errors with
respect to a finite subset of the spectrum will be considered. Denote
the rectangle spanned by two complex numbers $X,Y\in\mathbb{C}$ by
$\Omega(X,Y)$. The error between the true spectrum $\{\lambda_{n}^{\pm}\}_{n}$
and the numerical spectrum $\{\tilde{\lambda}_{j}\}_{j}$ with respect
to $\Omega(X,Y)$ can be measured by
\begin{align*}
e:=\max\Big\{ & \max_{\lambda_{n}^{\pm}\text{ s.t. }\lambda_{n}^{\pm}\in\Omega(X,Y)}\min_{\tilde{\lambda}_{j}\text{ s.t. }\tilde{\lambda}_{j}\in\Omega(X,Y)}|\lambda_{n}^{\pm}-\tilde{\lambda}_{j}|,\\
 & \max_{\tilde{\lambda}_{j}\text{ s.t. }\tilde{\lambda}_{j}\in\Omega(X,Y)}\min_{\lambda_{n}^{\pm}\text{ s.t. }\lambda_{n}^{\pm}\in\Omega(X,Y)}|\lambda_{n}^{\pm}-\tilde{\lambda}_{j}|\Big\}.
\end{align*}
Note that the first term in the outer maximum grows large if an algorithm
fails to approximate a part of the true spectrum within $\Omega(X,Y)$,
while the second term becomes large if an algorithm creates spurious
terms within $\Omega(X,Y)$ that have no correspondence in the true
spectrum. The second plot in Fig. \ref{fig:Runtimes_focusing} depicts
the error for $\Omega(-5+\im,5+5\im)$. That is, only the non-real
spectrum is considered. All four errors decrease approximately linearly
(i.e., doubling $D$ halves the error), but the errors of the fast
algorithms interestingly are lower than those of the standard algorithms
although the same discretizations are used. (This is only an apparent
discrepancy as the standard algorithms approximate the Lax operator
(\ref{eq:L}), while the fast algorithms approximate the monodromy
matrix (\ref{eq:monodromy-matrix}).) Fig. \ref{fig:spectra} illustrates
the different accuracies by comparing the exact and the numerical
spectra for $D=32$. The errors for $\Omega(-5+\im,5+5\im)$ are all
much higher (not shown) because all algorithms have problems with
the approximation of the eigenvalue at zero. However, the errors still
decrease linearly in that case and the relative performance of the
algorithms with respect to each other remains the same.

\subsection{The Defocusing Case\label{sub:Numerical-Example-Defocusing}}

\subsubsection{First Example: One-band Solution}

The next example is the $q(x,t_{0})=\frac{3}{2}\eu^{\im3x}$, where
the period is $\ell=2\pi$. This initial condition corresponds to
a one-band solution as derived in Example \ref{exa:Periodic-one-band-solution}
with $\lambda_{1}=-3$ and $\lambda_{2}=0$.The first plot in Fig.
\ref{fig:Runtimes_defocusing-one-band} depicts the per-sample runtime
of the fast NFT from Sec. \ref{sub:FNFT-Samp-Alg} with that of a
naive implementation where the monodromy matrix is evaluated directly
through (\ref{eq:iteration-AL}). In each case, $L=5$ bisection steps
have been carried out. The oversampling factor was $G=1$. The per-sample
runtime of the fast algorithm grows only very slowly with the number
of samples, while it grows linearly for the standard implementation.
The second plot in Fig. \ref{fig:Floquet-diagrams} shows the same
error as in Sec. \ref{sub:Numerical-Example-Focusing} where now $\{\tilde{\lambda}_{j}\}=\{-3,0\}$,
$X=-10$ and $Y=10$. We can see that the fast algorithm gives exactly
the same errors as the naive implementation. The first plot in Fig.
\ref{fig:Floquet-diagrams} finally shows the Floquet diagram (i.e.,
a plot of $\hat{\Delta}(z)$ where the scale is linear if $|\hat{\Delta}(z)|\le1$
and logarithmic otherwise) computed by the fast algorithm.

\subsubsection{Second Example: Gaussian Wavepacket}

\begin{figure*}[t]
\begin{centering}
\includegraphics[width=0.45\linewidth]{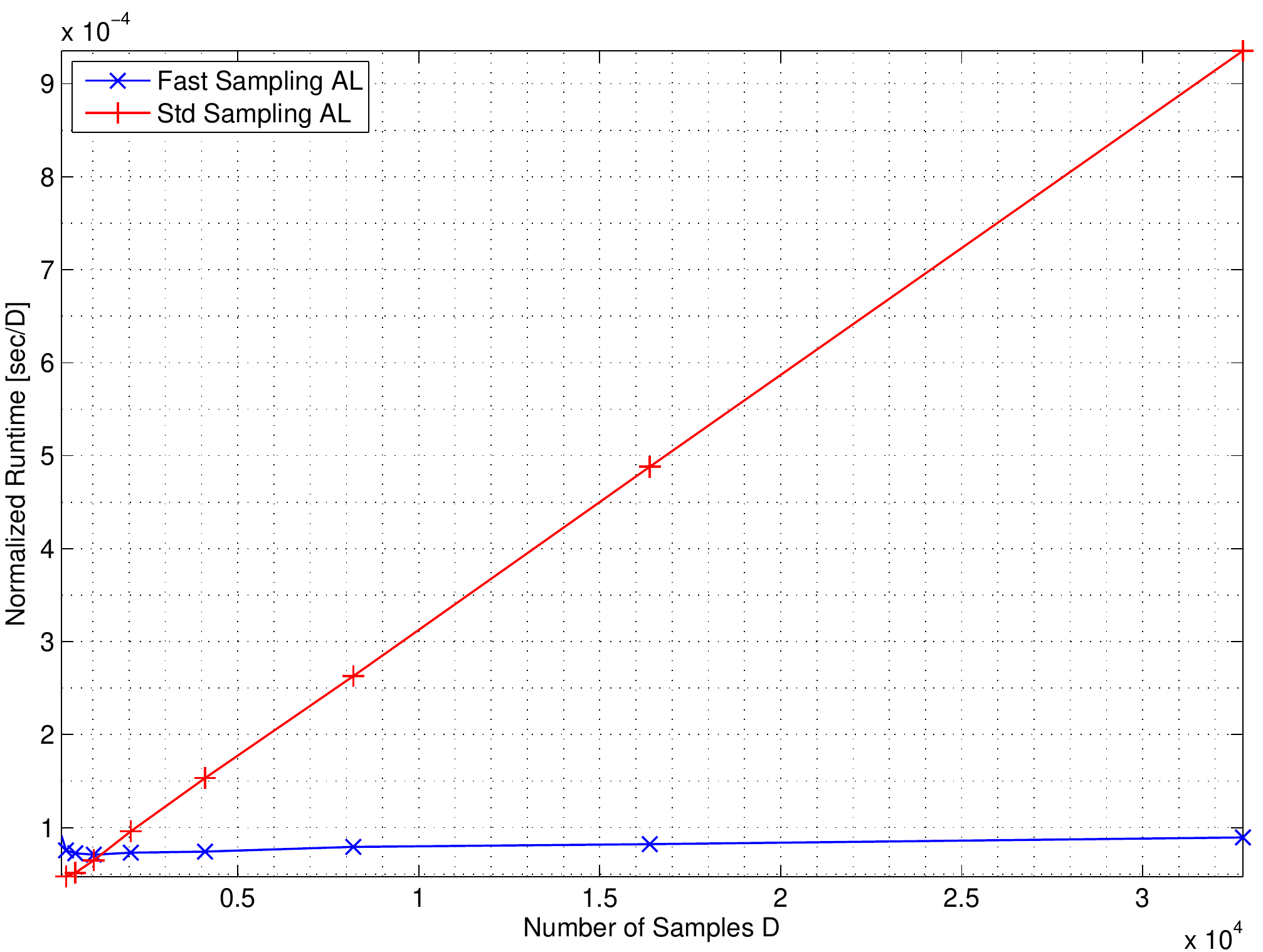}\qquad{}\includegraphics[width=0.45\linewidth]{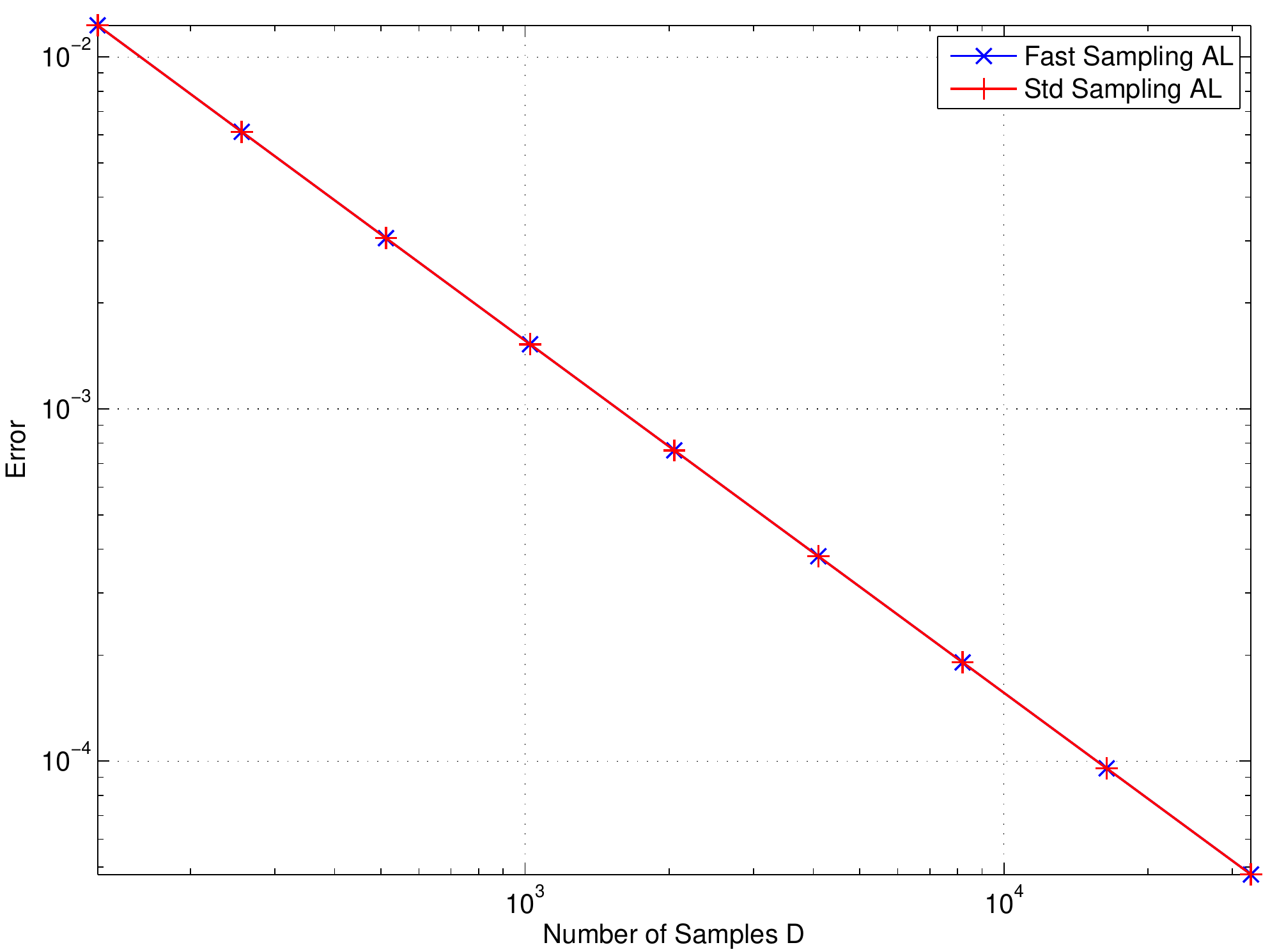}
\par\end{centering}

\centering{}\caption{\label{fig:Runtimes_defocusing-one-band}Defocusing one-band solution.
Left: Runtimes per sample, Right: Error.}
\end{figure*}

\begin{figure*}[t]
\begin{centering}
\includegraphics[width=0.45\linewidth]{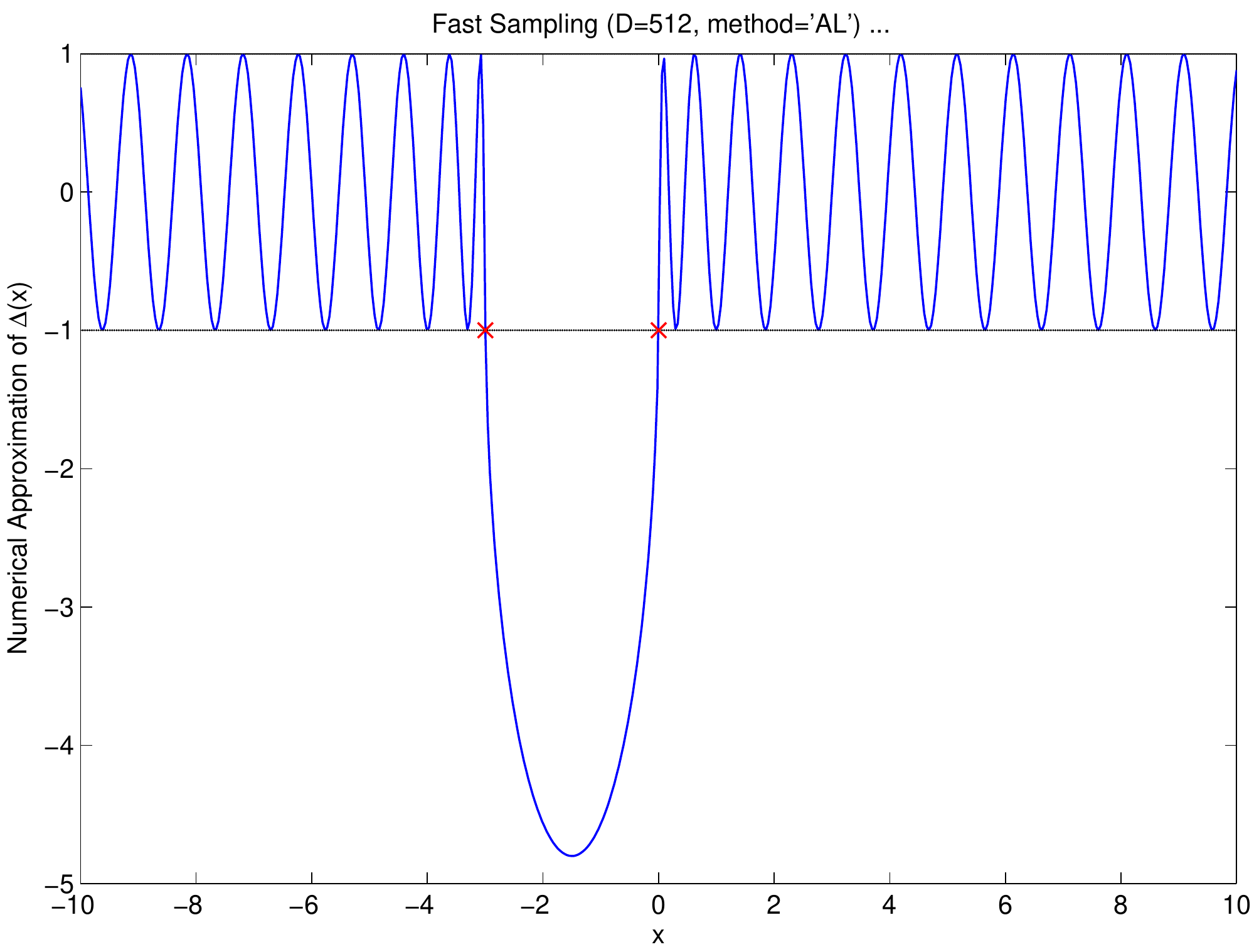}\qquad{}\includegraphics[width=0.45\linewidth]{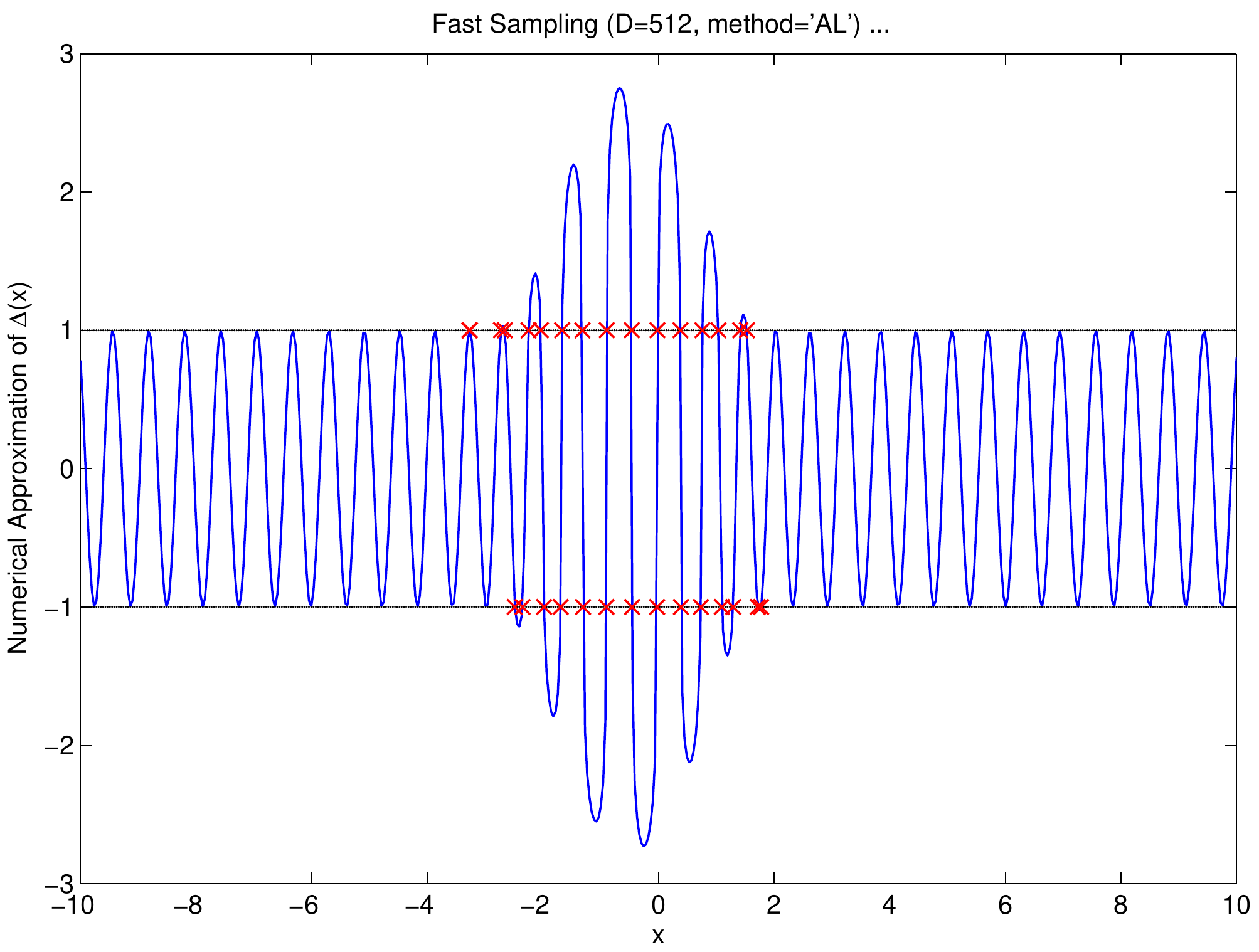}
\par\end{centering}

\centering{}\caption{\label{fig:Floquet-diagrams}Floquet diagram  (\textcolor{blue}{--})
and the main spectral points (\textcolor{red}{x}) for the one-band
solution (left) and the Gaussian wavepacket (right)}
\end{figure*}

The last example is $q(x,t_{0})=q_{0}\eu^{\im\mu x}\eu^{-\frac{x^{2}}{\sigma}}$,
which has been discussed in \cite{Osborne1993}. The parameters considered
here are $q_{0}=1.9$, $\mu=1$, $\sigma=2$, $\ell=10$. We do not
show the runtime plot because it is very similar to the one in Fig.
\ref{fig:Runtimes_defocusing-one-band}. The second plot in Fig. \ref{fig:Floquet-diagrams}
shows the Floquet diagram computed by the fast algorithm. While the
exact error cannot be quantified in this case because the analyical
NFT seems to be unknown, a comparison of the Floquet diagram to that
in \cite[Fig. 2b]{Osborne1993} confirms the result found by the fast
algorithm.

\section{Rapidly Decaying Signals \label{sec:Rapidly-Vanishing-Signals}}

The fast transforms derived in this paper can also be carried over
to rapidly decaying non-periodic signals on the line. Details on the
necessary modifications can be found in \cite{Wahls2013b}, where
a preliminary form of the eigenmethod in Sec. \ref{sec:FNSFT-Eigen}
has been presented for rapidly decaying signals. Alternatively, the
new algorithms for the periodic case may be used directly if the decaying
signal is truncated and extended periodically for a large enough period
$\ell>0$ \cite{Olivier2012}. This however requires some additional
transformations of the found spectra.

\section{Other Integrable Evolution Equations\label{sec:Other-Integrable-Evolution}}

The new fast NFTs presented in this paper have been developed for
signals that are governed by the NSE (\ref{eq:NSE}). However, the
approach extends to signals governed by other evolution equations
as well. In this section, the extension of our results will be discussed
for the \emph{Ablowitz-Kaup-Newell-Segur (AKNS) }Lax pair\emph{ }\cite{Ablowitz1974}.
(The Lax pair formalism has been introduced in Sec. \ref{sub:Lax-Pair-Formalism}.)
The authors did not investigate extensions to other Lax pairs so far,
although they feel that such extensions should be possible along lines
similar to those outlined below. Finally, please note that the proposed
extensions have not been investigated in numerical experiments so
far. Their numerical accuracy remains to be examined.

\subsection{The AKNS Lax Pair}

In Sec. \ref{sub:Lax-Pair-Formalism} it was mentioned that the NLS
(\ref{eq:NSE}) arises from the compatibility condition (\ref{eq:Lax-condition})
for certain Lax pairs. Ablowitz et al. \cite{Ablowitz1974} have established
that many other important evolution equations can be expressed through
the condition (\ref{eq:Lax-condition}) for Lax pairs with 
\begin{equation}
\mathbf{L}_{t_{0}}=\im\left[\begin{array}{cc}
\frac{d}{dx} & q(\cdot,t_{0})\\
r(\cdot,t_{0}) & -\frac{d}{dx}
\end{array}\right]\label{eq:AKNS}
\end{equation}
for suitably chosen signals $q$, $r$ and operator $\mathbf{B}$.
The standard examples of evolution equations that fall into this framework
(other than the NSE) are the following \cite[p. 258]{Ablowitz1974}:
\begin{itemize}
\item The\emph{ Korteweg-de Vries equation}: $r=1$, 
\[
\partial_{t}q+6q\partial_{x}q+\partial_{xxx}q=0.
\]

\item The \emph{modified KdV equation}: $r=\pm q$, 
\[
\partial_{t}q\pm6q^{2}\partial_{x}q+\partial_{xxx}q=0.
\]

\item The \emph{sine-Gordon equation}: $r=q=\frac{1}{2}\partial_{x}u$,
\[
\partial_{xt}u=\sin u.
\]

\item The \emph{sinh-Gordon equation}: $r=-q=\frac{1}{2}\partial_{x}u$,
\[
\partial_{xt}u=\sinh u.
\]

\end{itemize}
Either under additional transformations, or by considering matrix-valued
signals $\mathbf{q}$ and $\mathbf{r}$, many more equations can be
fit into the AKNS framework \cite{Manakov1974,Kaup1975,Jaulent1977,Gupta1979,Ishimori1982,Sakovich2005}.

\subsection{Finite-Band Solutions for the AKNS Lax Pair}

Tracy has presented finite-band solutions for the AKNS Lax pair in
\cite[Ch. 2.2]{Tracy1984a}. (An even more general case has been discussed
in \cite{Kamchatnov2001}.) One starts with exactly the same form
as in Sec. \ref{sec:Finite-Band-Solutions}. Then, one adds a second
set of $N-1$ auxiliary variables $\eta_{j}(x,t)$ and Riemann sheet
indices $\theta_{j}(x,t)\in\{\pm1\}$, respectively. The $\eta_{j}$
are governed by the differential equations 
\begin{align*}
\partial_{x}\eta_{j}= & \frac{2\im\theta_{j}\sqrt{\prod_{k=1}^{2N}(\eta_{j}-\lambda_{k}})}{\prod_{{m=1\atop m\ne j}}^{N-1}(\eta_{j}-\eta_{m})},\\
\partial_{t}\eta_{j}= & -2\left(\sum_{{m=1\atop m\ne j}}^{N-1}\eta_{m}-\frac{1}{2}\sum_{k=1}^{2N}\lambda_{k}\right)\partial_{x}\eta_{j}.
\end{align*}
The signal $r$ evolves according to 
\[
\partial_{x}\ln r=-2\im\left(\sum_{j=1}^{N-1}\eta_{j}-\frac{1}{2}\sum_{k=1}^{2N}\lambda_{k}\right).
\]
 The squared eigenfunction (\ref{eq:squared-eigenfunction-h}) has
to be replaced with 
\[
h_{z}(x,t)=\im r(x,t)\prod_{j=1}^{N-1}(z-\eta_{j}(x,t)).
\]
 Then, one has that $q$ and $r$ solve the system
\begin{align*}
\im\partial_{t}q+\partial_{xx}q+2q^{2}r & =0,\\
-\im\partial_{t}r+\partial_{xx}r+2r^{2}q & =0
\end{align*}
if and only if (\ref{eq:squared-eigenfunction-f}) is a polynomial
\cite[Thm. 2.1]{Tracy1984a}.

\subsection{Fast NFTs for the AKNS Lax Pair}

The algorithms presented in this paper can easily be extended to general
AKNS Lax pairs. All results except Lemma \ref{lem:monodromy-symmetry-like-props},
Lemma \ref{lem:main-spectrum-symmetry}, and Sec. \ref{sub:Ma-Ablowitz}
 on the computation of the scattering data in Sec. \ref{sec:The-NFT}
carry over to the general AKNS case if the operator $\mathbf{L}_{t_{0}}$
in (\ref{eq:L}) is replaced with (\ref{eq:AKNS}) and Eq. (\ref{eq:Lv-lamv-as-DE})
is updated accordingly. The scattering data has to be extended by
the initial conditions $\eta_{j}(x_{0},t_{0})$, which  turn out to
be the roots of $\left[\mathbf{M}_{x_{0},t_{0}}(z)\right]_{2,1}$.

Numerically, only minor changes are necessary in the development of
Sec. \ref{sec:Rational-Approximations-of-M} in order to adapt the
rational approximations of the monodromy matrix if the operator $\mathbf{L}_{t_{0}}$
is changed. Basically, only the matrix $\mathbf{P}_{z}$ defined in
(\ref{eq:P}) has to be changed such that the eigenproblem $\mathbf{L}_{t_{0}}\mathbf{v}=z\mathbf{v}$
is again equivalent to $\frac{d}{dx}\mathbf{v}=\mathbf{P}_{z}\mathbf{v}$.
Using (\ref{eq:AKNS}), one finds that $\mathbf{P}_{z}$ is given
by 
\begin{equation}
\frac{d}{dx}\mathbf{v}=\left[\begin{array}{cc}
-\im z & -q(\cdot,t_{0})\\
r(\cdot,t_{0}) & \im z
\end{array}\right]\mathbf{v}=:\mathbf{P}_{z}\mathbf{v}.\label{eq:general-Pz}
\end{equation}
Of course, consecutive terms that involve $\mathbf{P}_{z}$ have to
be reevaluated. Then, our proposed fast algorithms in Secs. \ref{sub:FNSFT-Eigen-Alg}
and \ref{sub:FNFT-Samp-Alg} can be run as before. The methods used
to find the $\mu_{j}(x_{0},t_{0})$ can be used to find the $\eta_{j}(x_{0},t_{0})$
as well.

\section{Conclusion\label{sec:Conclusions}}

In this paper, two fast numerical nonlinear Fourier transforms for
the periodic nonlinear Schr\"odinger equation have been proposed.
The first algorithm has a complexity of $O(D^{2})$ flops, where $D$
denotes the number of sample points. The second algorithm applies
only to the defocusing nonlinear Schr\"odinger equation, but its
complexity is only $O(D\log^{2}D)$ flops. In both cases, this is
about an order of magnitude better than what other comparable algorithms
achieve so far. The feasibility of the fast transforms has been demonstrated
in several numerical examples. Extensions to other cases have been
discussed as well.

\bibliographystyle{IEEEtran}
\bibliography{library}

\section*{}

\end{document}